\documentclass{aa}  
\usepackage{graphicx}
\usepackage{booktabs}
\usepackage{longtable}
\usepackage{scalerel}
\newcommand{\smallsub}[2] {#1^{}_{\scaleto{\mathrm{#2}}{4pt}}}

\newcommand{\smallsubTilde}[2] {\tilde{#1}^{}_{\scaleto{\mathrm{#2}}{4pt}}}
\newcommand{\as}{\,\mathrm{arcsec}}
\newcommand{\am}{\,\mathrm{arcmin}}
\usepackage{txfonts}
\usepackage{hyperref}
\usepackage{xcolor}
\hypersetup{
    colorlinks,
    linkcolor={blue!100!black},
    citecolor={blue!100!black},
    urlcolor={blue!100!black}
}
\makeatletter
\renewcommand*\aa@pageof{, page \thepage{} of \pageref*{LastPage}}
\makeatother
\begin{document}

   \title{Gaia--TESS synergy: \\ Improving the identification of transit candidates}

\author{
Aviad Panahi \inst{1} \and 
Tsevi Mazeh \inst{1} \and 
Shay Zucker\inst{2} \and 
David W. Latham \inst{3} \and
Karen A. Collins \inst{3} \and
Lorenzo Rimoldini\inst{4} \and
Dafydd Wyn Evans\inst{5} \and 
Laurent Eyer\inst{4}
}

\institute{
School of Physics and Astronomy, Raymond and Beverly Sackler Faculty of Exact Sciences, Tel Aviv University, Tel Aviv 6997801, Israel
\and
    Porter School of the Environment and Earth Sciences, Raymond and Beverly Sackler Faculty of Exact Sciences, Tel Aviv University, Tel Aviv 6997801, Israel
\and
    Center for Astrophysics, Harvard \& Smithsonian, 60 Garden Street, Cambridge, MA 02138, USA
\and 
    Department of Astronomy, University of Geneva, Chemin Pegasi 51,
    1290 Versoix, Switzerland
\and
    Institute of Astronomy, University of Cambridge, Madingley Road,
    Cambridge CB3 0HA, UK
}

   \date{Received x x, x; accepted x x, x}

\abstract
{The TESS team periodically issues a new list of transiting exoplanet candidates based on the analysis of the accumulating light curves obtained by the satellite. The list includes the estimated epochs, periods, and durations of the potential transits. As the point spread function (PSF) of TESS is relatively wide, follow-up photometric observations at higher spatial resolution are required in order to exclude apparent transits that are actually blended background eclipsing binaries (BEBs).}
{The Gaia space mission, with its growing database of epoch photometry and high angular resolution, enables the production of distinct light curves for all sources included in the TESS PSF, up to the limiting magnitude of Gaia.
This paper reports the results of an ongoing {Gaia}-TESS collaboration that uses the {Gaia} photometry to facilitate the identification of BEB candidates and even to confirm on-target candidates in some cases.}
{We inspected the Gaia photometry of the individual sources included in the TESS PSF, searching for periodic dimming events compatible with their ephemerides and uncertainties, as published by TESS.
The performance of the search depends mainly on the number of Gaia measurements during transit and their precision.}
{Since February 2021, the collaboration has been able to confirm $126$ on-target candidates and exclude $124$ as BEBs. 
Since June 2021, when our search methodology matured, we have been able to identify on the order of $5\%$ as on-target candidates and another $5\%$ as BEBs. }
{This synergistic approach is combining the complementary capabilities of two of the astronomical space missions of NASA and ESA. It serves to optimize the process of detecting new planets by making better use of the  resources of the astronomical community.
}

   \maketitle
%

\section{Introduction}
\label{sec:intro}

The Transiting Exoplanet Survey Satellite \citep[TESS;][]{Ricker2015} is systematically scanning the sky in search of transiting exoplanets, producing hundreds of thousands of photometric light curves.
TESS collects its data using four wide-field cameras with a relatively wide point spread function (PSF), namely
on the order of $1\am$ for bright stars \citep{Sullivan15}.
Faint unknown background eclipsing binary stars (BEBs) residing within the PSF of the TESS target star might therefore cause photometric modulations that would appear as planetary transits \cite[e.g.,][]{Brown03,Cameron2012}. For this reason, photometric follow-up observations with high spatial resolution are required for the TESS objects of interest (TOIs) and their nearby stars to determine the actual source of the transit-like signal \cite[e.g.,][]{Astudillo20, Waalkes21, Gan21}. This is usually done by ground-based seeing-limited photometric observations that can be used to differentiate between true and false-positive detections \citep[e.g.,][]{Deeg09, Collins18, Lendl20}.

To identify the source of the modulations seen by {TESS}, one can also use the Gaia photometric data that have been accumulating since the launch of the European space mission in 2013 \citep{Prusti2016}.
Although the main objective of the Gaia mission is high-accuracy astrometry of more than a billion objects \citep{GaiaEDR3}, {Gaia} also produces photometric measurements in its wide $G$ 
band, which mostly covers the visible wavelength range \citep{Jordi2010}.

By itself, Gaia photometry is not optimized for detecting transiting planets, as the satellite scans the sky in an irregular low-cadence fashion, with typically a dozen observations per year \citep{Prusti2016}. On the other hand, TESS (during its prime mission) has been continuously monitoring the sky for about $27.5$ days per sector, with a high cadence of 2 minutes for bright stars and 30 minutes for faint ones \citep{Ricker2015}. 
Nevertheless, thanks to the large number of stars sampled by Gaia and the photometric precision that can be on the order of a few millimagnitudes (mmag) or less \citep{GaiaEDR3},
two new transiting hot-Jupiters have already been discovered using Gaia photometry and validated via radial velocities \citep{GaiaPlanets_ADS}. More candidates were published in Gaia DR3 \citep{Eyer22}.

The high angular resolution of Gaia currently enables it to safely resolve sources with a separation above $\sim0.7\as$, and it is expected to even improve in future data releases \citep{GaiaEDR3_catval}.
In the case where a BEB is the origin of an apparent transit detected by TESS, Gaia may have already observed the eclipse photometrically. In this case, as Gaia is observing the undiluted and deeper transit, it can identify this star as the culprit of the TESS false positive. However, this is only possible if Gaia photometry has indeed sampled the eclipse events of the background binary. Short-period binaries might also display an ellipsoidal variability \citep[e.g.,][]{Morris93,Faigler11} that can be detected by Gaia if the companion is sufficiently massive. If the secondary is sufficiently luminous, we might also see a secondary eclipse. On the other hand, the object eclipsing the background star might actually be a planet, shifting the planet-host candidacy to the nearby star.
In case where the observed transit is due to the brightest star within the PSF, and the transit is deep enough to be discerned in the Gaia photometry, we can confirm the authenticity of the TOI as a transiting planet candidate ("on-target confirmation").

Assessing the source of the transit signal has a time-sensitive nature, as successful identifications can save valuable observational resources. Joining the forces of the two space missions became imperative.
We therefore initiated a collaboration between Gaia Data Processing and Analysis Consortium \cite[DPAC;][]{Prusti2016}
and 
SubGroup~1 of the TESS Follow-up Observing Program Working Group
\citep[SG1;][]{Col2019}, liaised by the Tel Aviv University (TAU) group (which is part of DPAC), whose goal is to examine the TESS candidates with the help of the accumulating Gaia data. The TAU group examines the unpublished Gaia photometry to identify genuine and false candidates and shares those identifications with SG1 shortly after the release of the candidates. This collaboration started in February 2021, around the time TESS announced TOI-$2400$. 
The detection capability of the search gradually improved until June 2021, when TESS released TOI-$3504$.

This paper presents the details of the {Gaia}-{TESS} collaboration. 
We present the search methodology and list $124$ BEBs, $124$ on-target confirmations, and two planets in wide binary systems.
Section~\ref{sec_algo} describes the search method for cases in which Gaia had measured dimming events in the TOIs or their neighboring stars and the initial assessment of whether these cases were on target. Section~\ref{sec:results} presents the current results of the collaboration, including examples of on-target confirmations and identified BEBs.  Section~\ref{sec:performance} presents an early assessment of the performance of the procedure and Sect.~\ref{sec:discussion} discusses the significance of this collaboration and its future.

\section{Search procedure}
\label{sec_algo}
For each TOI, the procedure starts by retrieving the Gaia $G$-band photometry of each source within $1\am$ of the position reported for the TOI; such sources are referred to as PSF neighbors hereafter. 
This radius was chosen in order to roughly accommodate the angular size of 3 TESS pixels, or an aperture area of $\pi \cdot 3^2 \approx 30$ pixels, which should include most of the accumulated light (see discussion below and Fig.~13 in \cite{Sullivan15}). 
In each light curve, we remove points that are brighter than the median magnitude by more than  three times the median absolute deviation (MAD) of the photometric time series.
After removing these points, only sources that have a minimum number of Gaia measurements are considered. Our most limiting criterion is the minimum number of in-transit points. We chose this number to be two, because one point is insufficient to constrain the periodicity and three would be too strict considering the sparsity of the Gaia photometry. We chose the minimum total number of points to be $14$, somewhat arbitrarily, in an attempt to avoid dealing with small number statistics for the out-of-transit points.

Next, we explore the parameter space allowed by the uncertainties of the TESS parameters by applying a modified version of the box-fitting least square algorithm \citep[BLS;][]{Kovacs2002} to the Gaia data, scanning a small region in the two-dimensional space around the TOI nominal period ($\smallsub{P}{TOI}$) and transit duration ($\smallsub{d}{TOI}$), as reported by the TOI alert. The goal is to find the parameter values that produce the highest BLS signal residue ($\mathcal{SR}$)\footnote{The $\mathcal{SR}$ is related to the sum of squared residuals in the least-squares fit of the attempted box-shaped model.} statistic \citep{Kovacs2002}, while having at least two measurements in-transit.
The search is restricted to a range of $\pm 3\Delta$ of the period and duration, where $\Delta$ is the respective uncertainty as reported by TESS. 

We do not restrict the mid-transit time $T$ at this stage and scan the whole range of phases in steps of one-tenth of the nominal transit duration.
To incorporate the prior knowledge of the transit events, we apply Gaussian priors for the transit period, mid-transit time, and duration,
resulting in a 
modified signal residue ($\mathcal{MSR}$), 
%
\begin{equation}
    \mathcal{MSR} = \mathcal{SR} - \left( \dfrac{P - \smallsub{P}{TOI}}{\Delta \smallsub{P}{TOI}} \right)^2 - 
                                   \left( \dfrac{T - \smallsub{T}{TOI}}{\Delta \smallsub{T}{extended}} \right)^2 - 
                                   \left( \dfrac{d - \smallsub{d}{TOI}}{\Delta \smallsub{d}{TOI}} \right)^2 \,.
\end{equation}
%
The prior for $T$ includes an extended uncertainty ---taking into account the reported uncertainty for $\smallsub{T}{TOI}$ and the accumulated uncertainty of the period--- due to the number of cycles between the Gaia and TESS measurements:
\begin{equation}
    \Delta \smallsub{T}{extended} = \sqrt{\left( \Delta \smallsub{T}{TOI} \right)^2 
                                        + \left( \dfrac{\smallsub{\tilde{t}}{TESS}-\smallsub{\tilde{t}}{Gaia}}{\smallsub{P}{TOI}} \Delta \smallsub{P}{TOI} \right)^2 }\,                                        
                                        ,
\end{equation}
where $\tilde{t}$ is the average observation time of the candidate, for each of the two space missions. 
Using the extended uncertainty for $T$ lowers the weight of the mid-transit time prior, thus allowing the phase to shift and appear as a variation in the transit time.

The algorithm then selects the light curve that yields the highest $\mathcal{MSR}$, using the best-fitting parameters to derive the transit depth $\delta$, as seen by Gaia, and a merit function defined as the transit signal-to-noise ratio:
\begin{equation} \label{eq:SNR}
    \mathcal{S/N}_\mathrm{T}=\dfrac{\delta}{\smallsub{\sigma}{OOT}} \sqrt{\smallsub{N}{IT}} \, .
\end{equation}
Here $\smallsub{\sigma}{OOT}$ is the standard deviation of the out-of-transit (OOT) measurements (as a proxy to the inherent variability of the light curve, ignoring the transits) and $\smallsub{N}{IT}$ is the number of in-transit (IT) 
{Gaia} measurements. 
We found most stars with $\mathcal{S/N}^{}_\mathrm{T}>7$ to be good candidates either for confirming or refuting the TOI candidacy and only very few with $\mathcal{S/N}^{}_\mathrm{T}<7$.

We proceed to inspect the candidates individually and group them into on-target and BEB candidates based on their angular distance from the TOI location.
For on-target candidates, we make sure that the star is the brightest or one of the brightest stars in the TESS PSF, and that the transit depths of Gaia and TESS are similar.
For BEB candidates, we check that the observed depth in the Gaia light curve is compatible with the depth reported by TESS, after correcting for dilution, and make sure the Gaia light curve of the target star shows no significant variability.
We then proceed to visually examine the candidates to finally identify the BEB and on-target cases.
\section{Results}
\label{sec:results}

\subsection{Overview}
The collaboration has so far gone  through two phases.
In the first preliminary phase (henceforth Phase~I), during February-May 2021, the search procedure was still evolving, while producing partial results: $52$ BEBs, $29$ on-target confirmations, and one planet in a wide binary system. Since the second phase (Phase~II) started in June 2021, a regular search has been performed. After examining $1600$ TOIs in Phase~II ($3504$\,--\,$5103$), the search yielded $72$ BEBs, $95$ on-target confirmations, and one planet in a binary system. As mentioned in Sect.~\ref{sec_algo}, we included in our analysis only Gaia light curves with a minimum number of $14$ measurements, after removing outliers.

We list the results of the two phases in Tables~\ref{tab:rest_params_conf}, \ref{tab:rest_params_bebs}, \ref{tab:group_params_conf}, and \ref{tab:group_params_bebs}. 
The tables include the period and depth of the released TOIs, the number of measurements in the Gaia light curve, $N$, the number of in-transit measurements, $N^{}_\mathrm{IT}$, and the derived depth of the Gaia transit (or eclipse in case of a BEB). All transit depths are given in parts per thousand (ppt). The depths in the Gaia photometry are calculated based on the box-shaped model of the BLS algorithm. The small number of points and the precision available do not allow us to fit a more accurate shape to the transit, which includes the limb-darkening effect
\citep[e.g.,][]{Mandel2002, Hippke2019}.

In total, since February 2021, the search has yielded $124$ BEBs, $124$ on-target confirmations, and two planets in wide binary systems. We were also able to confirm a few dozen cases of known planets, but for conciseness we chose to exclude them from the present discussion.

\subsection{Examples}
\label{sec:examples}

We present the details of five specific cases that demonstrate the various types of discoveries made by the collaboration: two on-target cases confirmed by Gaia,
for which the transits reported by TESS originate from the TOI target stars themselves, 
two BEBs, and one on-target transit in a wide binary system. 

We obtained the TESS full-frame-image (FFI) photometry using the \texttt{Python} libraries \texttt{Lightkurve} \citep{Lightkurve}, \texttt{Eleanor} \citep{Eleanor}, and \texttt{TESScut.MAST} \citep{Astrocut}. We used the \texttt{flatten()} method of the \texttt{Lightkurve} library to detrend the data.
In the following figures, the Gaia and TESS light curves are phase-folded with the period and mid-transit time obtained by our analysis.


\subsubsection{On-target confirmations}
\begin{figure}
\includegraphics[width=1\linewidth]{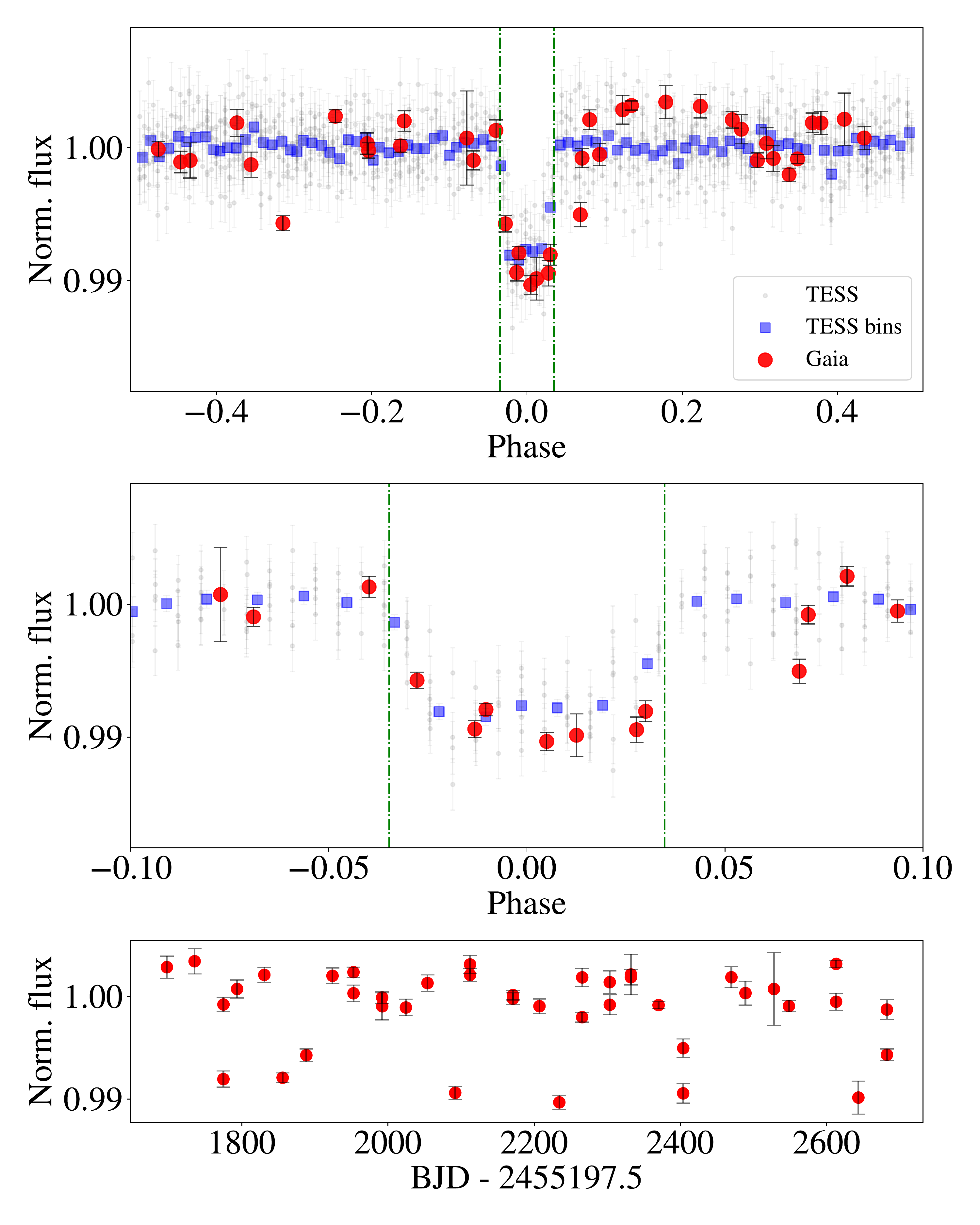}
\caption{On-target confirmation of TOI-$4034$.
\textit{Top:}
Gaia and {TESS} light curves of TOI-$4034$, phase folded with a period of $P=1.80209590$ days. 
\textit{Middle:} Zoomed-in view of the transit. 
\textit{Bottom:} Unfolded Gaia light curve.
The Gaia light curves are marked by red circles, the TESS light curves are marked by light-gray squares, and the bins by blue squares. The duration of the transit, as reported by TESS, is marked by two vertical
 green dashed lines.}
\label{fig:Conf_4034} 
\end{figure}

\begin{figure}
\includegraphics[width=1\linewidth]{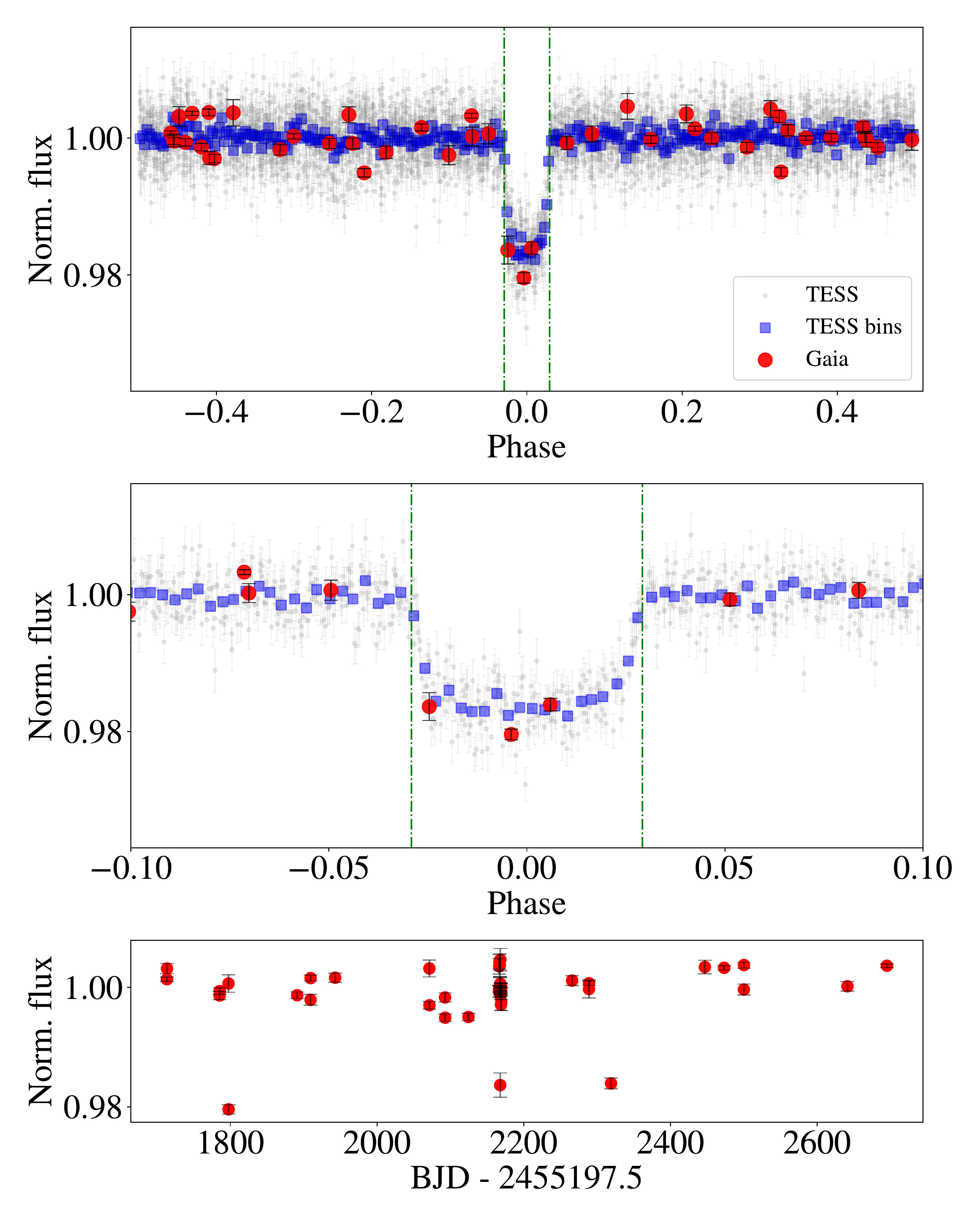}
\caption{On-target confirmation of TOI-$4293$.
\textit{Top:} 
Gaia and {TESS} light curves of TOI-$4293$, phase folded with a period of $P=1.62421090$ days.
\textit{Middle:} Zoomed-in view of the transit. 
\textit{Bottom:} Unfolded Gaia light curve.
See Fig.~\ref{fig:Conf_4034} for details of the lines and symbols.
}
        \label{fig:Conf_4293} 
\end{figure}

Figures~\ref{fig:Conf_4034} and \ref{fig:Conf_4293} show two on-target confirmations, with little or no dilution in the TESS light curves.
We note that even two or three Gaia measurements in transit can be sufficient to confirm the transit events.

\subsubsection{Background eclipsing binaries}
\label{sec:BEB}

\begin{figure}
\includegraphics[width=1\linewidth]{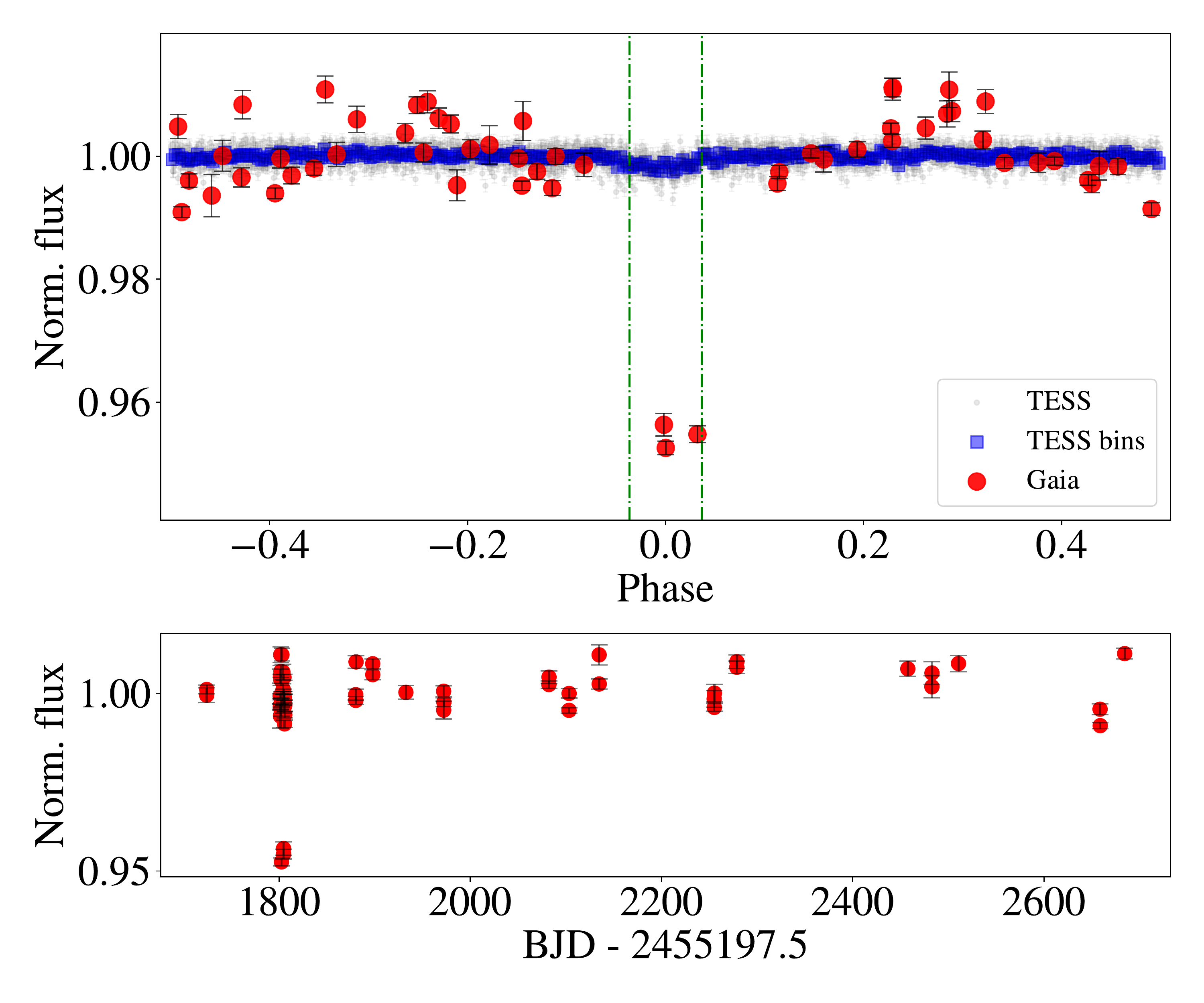}
\caption{BEB found near TOI-$4216$.
\textit{Top:} 
TESS light curve of TOI-$4216$, phase folded with a period of $P=2.18168090$ days, together with the Gaia light curve of its BEB PSF neighbor.
We note the clear ellipsoidal effect often seen in close-by binaries (see text). \textit{Bottom:} Unfolded Gaia light curve.
See Fig.~\ref{fig:Conf_4034} for details of the lines and symbols.}
        \label{fig:BEB_4216} 
\end{figure}

\begin{figure}
\includegraphics[width=1\linewidth]{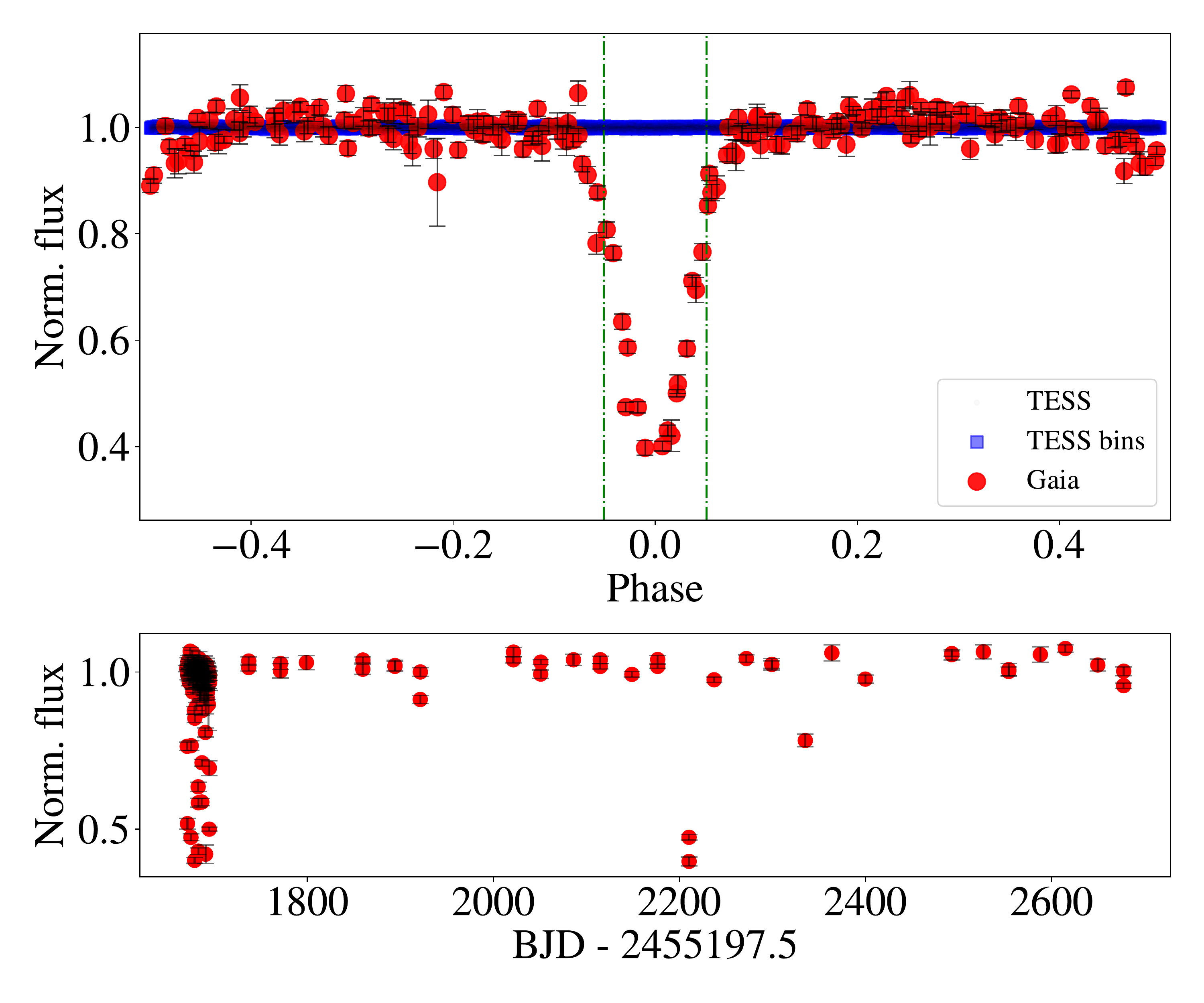}
\caption{BEB found near TOI-$4346$.
\textit{Top:}
TESS light curve of TOI-$4346$, phase folded with a period of $P=3.907574060$ days, together with the Gaia light curve of its BEB PSF neighbor. We note the clear secondary eclipse of the BEB centered at phase $0.5$, indicating a circular orbit.
The {TESS} light curve does not show the transit due to the vertical scale, determined by the deep eclipse.
\textit{Bottom:} Unfolded Gaia light curve. See Fig.~\ref{fig:Conf_4034} for details of the lines and symbols.}
        \label{fig:BEB_4346} 
\end{figure}

Figures~\ref{fig:BEB_4216} and \ref{fig:BEB_4346} show two BEBs identified by {Gaia} photometry, both found less than $1\am$ away from the position reported for their TOI. In both cases, the periodicity of the BEB star is consistent with the parameters of the TESS transit.
The Gaia depths are larger, because the angular resolution of Gaia enables it to see the undiluted eclipse. The Gaia light curve of the TESS target star showed no significant variability. Figure~\ref{fig:BEB_4216} displays an ellipsoidal modulation, and Fig.~\ref{fig:BEB_4346} shows a secondary eclipse, both features point to the stellar nature of the companion.

\subsubsection{Planet in a binary system}
\begin{figure}
        \includegraphics[width=1\linewidth]{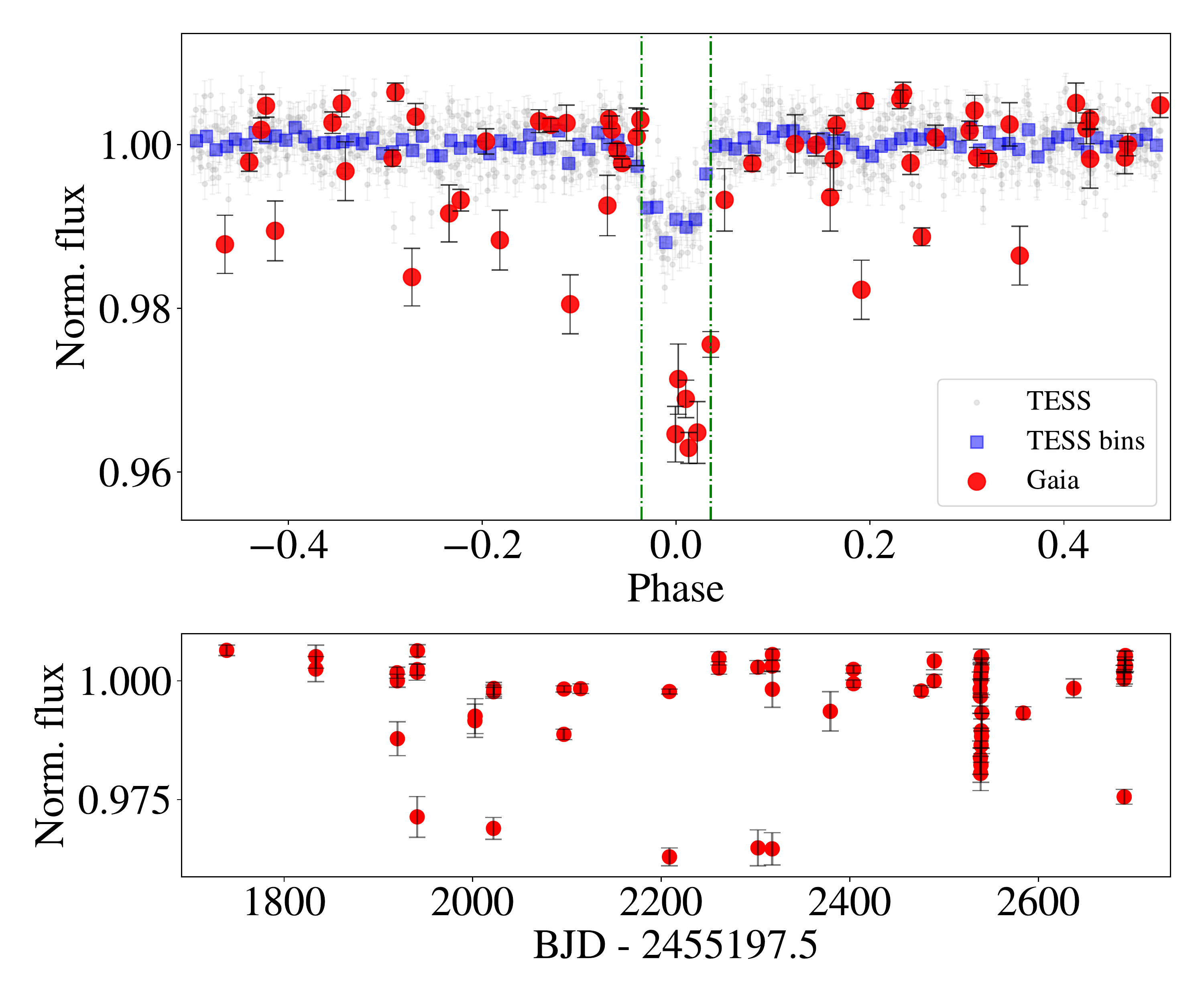}
\caption{On-target confirmation of a planet in a binary system, TOI-$3722$.
\textit{Top:} 
Gaia and {TESS} light curves of TOI-$3722$, phase folded with a period of $P=1.078945850$ days. The transit in the TESS light curve is diluted by the light of a neighboring star, which is suspected to form a binary system with the host star.
\textit{Bottom:} Unfolded Gaia light curve. See Fig.~\ref{fig:Conf_4034} for details of the lines and symbols.}
        \label{fig:BP_3722} 
\end{figure}

Figure~\ref{fig:BP_3722} shows the transit of TOI-$3722$, suggesting an on-target identification. However, the significantly deeper transit observed by {Gaia} indicates a nearby star unresolved by {TESS}. Indeed, Gaia~EDR3 lists two stars separated by $\sim1.2\as$ at the TOI location: 490902499803340800 (A), exhibiting the transit in Gaia data, and 490902465447630976 (B) which does not exhibit any transit-like features. Based
on the proximity between their locations on the sky, their parallax, and
proper motion (Table~\ref{tab:BinaryParams}), we propose that the two stars, with a projected separation of $830 \pm 10$ AU, form a wide binary.
Using the criteria listed by \citet{ElBadryetal2021}, the spatial parameters of the two stars are consistent with a bound binary system, although \citeauthor{ElBadryetal2021} do not include the pair in their final catalog.
\begin{table*}
\renewcommand{\arraystretch}{1.2}
    \centering
    \caption{TOI-$3722$: {Gaia} EDR3 parameters of the two stars suspected to be in a binary system.}
    \begin{tabular}{ c | c c c c c c c}
        Star & $\alpha$~(J2016.0) & $\delta$~(J2016.0) & $\mu_{\alpha\ast}$ & $\mu_{\delta}$ & $\varpi$ & $G$ & $G_\mathrm{BP} - G_\mathrm{RP}$ \\
         & h:m:s & d:m:s & $\mathrm{mas}~\mathrm{yr}^{-1}$ & $\mathrm{mas}~\mathrm{yr}^{-1}$ & $\mathrm{mas}$ & mag & mag \\
         \hline
        A & 3:21:18.05 & +63:43:56.75 & $3.361 \pm 0.021$ & $-3.038 \pm 0.029$ & $1.403 \pm 0.030$ & $14.6773 \pm 0.0028$ & $1.204 \pm 0.062$ \\
        B & 3:21:17.89 & +63:43:57.37 & $3.602 \pm 0.018$ & $-2.542 \pm 0.022$ & $1.449 \pm 0.026$ & $14.6342 \pm 0.0028$ & $1.164 \pm 0.027$\\
    \end{tabular}
    \label{tab:BinaryParams}
\end{table*}


\section{Performance} \label{sec:performance}
\subsection{Discovery rate} \label{sec_yield}

\begin{figure}[ht]
        \includegraphics[width=1\linewidth]{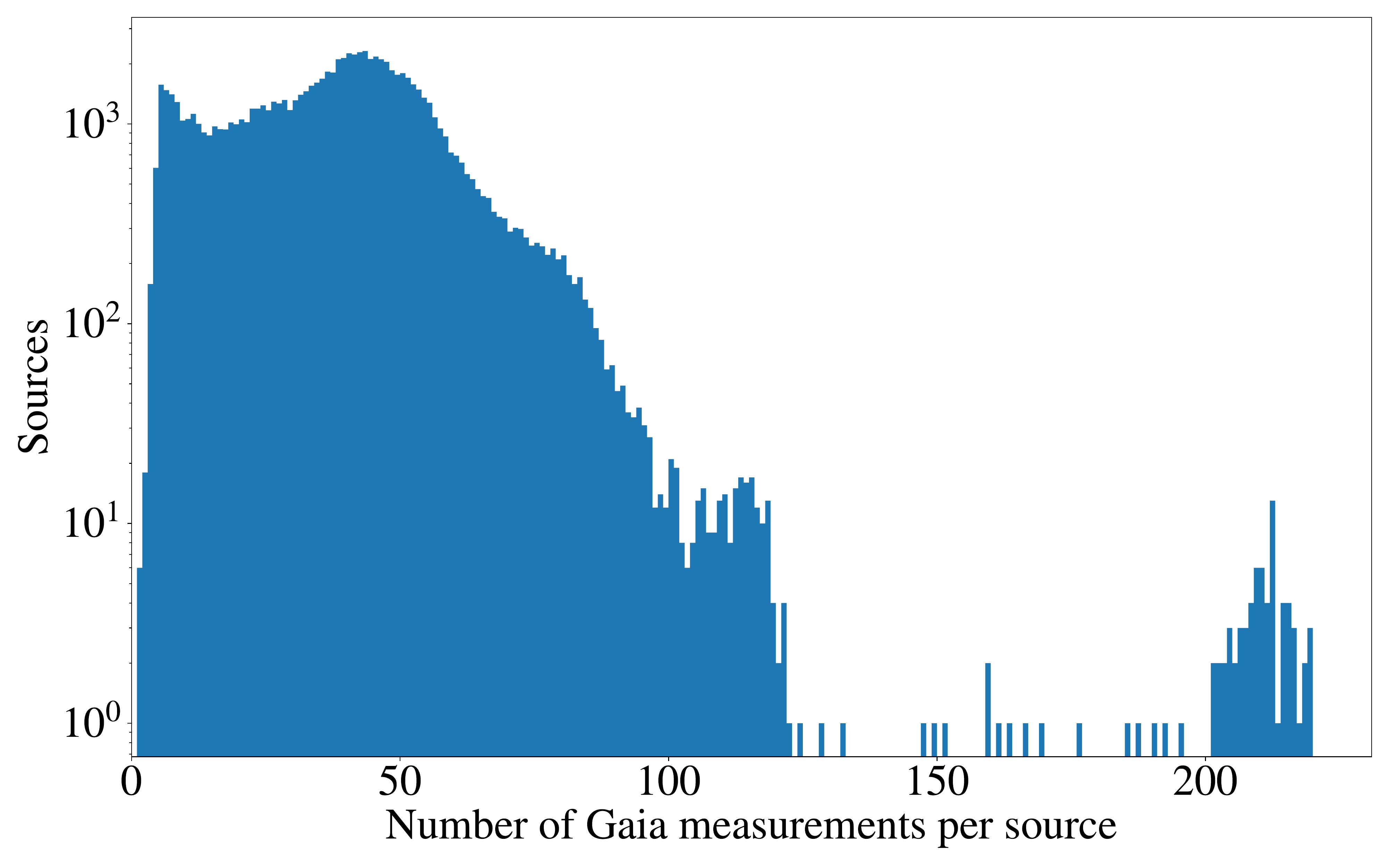}
        \caption{Phase~II: Number of Gaia measurements per star for all $89\,460$ objects  around the $1600$ TOIs in Phase~II. We note the cluster of stars with more than $200$ measurements, which is due to the scanning law of Gaia at the beginning of the mission (see text).}
        \label{fig:General_Histogram} 
\end{figure}

As explained above, 
we consider on-target confirmations or BEB detections only if there are at least two points in transit. Therefore, when estimating the expected rate of confirmations or detections, it is important to consider how many Gaia measurements per star were available at the time of the search.
We chose to use the sample of $1600$ Phase~II TESS candidates (TOIs $3504$\,--\,$5103$) for this purpose.
In the Gaia database, we find $89\,460$ PSF neighbors of the $1600$ TOIs, $77\,049$ of them having at least $14$ Gaia measurements, which is the lower limit we chose for our current analysis (Fig.~\ref{fig:General_Histogram}).
There are about $200$ measurements for one cluster of sources, which is due to stars in the ecliptic poles, which Gaia sampled extensively during the first $28$ days of its operation \citep{Prusti2016}.

For a given star, the expected number of in-transit measurements, $\smallsub{\mathcal{N}}{IT}$, can be estimated by multiplying the number of its Gaia measurements, $N$, by the fractional duration of the transit (duration divided by the period):
%
\begin{equation} \label{eq:NIT}
    \smallsub{\mathcal{N}}{IT} = N \frac{\smallsub{d}{TOI}}{\smallsub{P}{TOI}} \, .
\end{equation}

When considering the discovery rate of on-target confirmations, it is important to note that the published transit depths of TOIs can be quite small. As the photometric precision of Gaia depends on the target brightness in a complicated manner \citep{GaiaEDR3_Phot}, we consider only cases that are detectable by Gaia in terms of their expected $\mathcal{S/N}_\mathrm{T}$. We therefore only consider cases for which $\mathcal{S/N}_\mathrm{T} > 5$. This leaves $148$ cases out of the $1600$ from Phase~II, of which only $85$ had at least two measurements expected in transit.\footnote{The expected number of in-transit measurements is rounded to the nearest integer, and therefore the actual criterion we used in the estimate is $\smallsub{\mathcal{N}}{IT} > 1.5$.} 

BEBs on the other hand are expected to display substantially deeper eclipses, easily detectable by {Gaia}.
However, the number of measurements of the star, $N$, for BEBs can only be estimated, because the identity of the eclipsing binary is not known in advance. We therefore use $\tilde{\mathcal{N}}$  instead, which is the median number of Gaia measurements of the PSF neighbors of each TOI, and the corresponding  expected number of measurements in transit,  $\smallsubTilde{\mathcal{N}}{IT}$.

Within the sample of $1600$ TOIs in Phase~II, there are $562$ TOIs with $\smallsubTilde{\mathcal{N}}{IT}\geq 1.5$.
According to \citet[their Fig.~2]{Sullivan15}, the estimated fraction of BEB false positive TOIs 
is $f_\mathrm{BEB} = 0.15 \pm 0.08$, corresponding to $250 \pm 125$ BEBs in the Phase~II sample and $87\pm43$ in the restricted sample of $562$ TOIs.

For on-target detections, the fraction given by \citeauthor{Sullivan15} is $f_\mathrm{on-target} = 0.84 \pm 0.12$, corresponding to $1350 \pm 190$ in the Phase~II sample and $73 \pm 10$ in the restricted sample of $85$ TOIs.
Comparing these estimates with the $72$ detected BEBs (vs.~$87$ estimated) and $87$ on-target confirmations with $\mathcal{S/N}_\mathrm{T} > 5$ (vs.~$73$ estimated), we suggest that the search for both BEBs and on-target confirmations is performing as expected, with a detection rate of $\sim5\%$ for each case. The main limiting factor is the number of {Gaia} measurements. 
\subsection{Transit depths} \label{sec:transit_depths}
\begin{figure}
\includegraphics[width=1\linewidth]{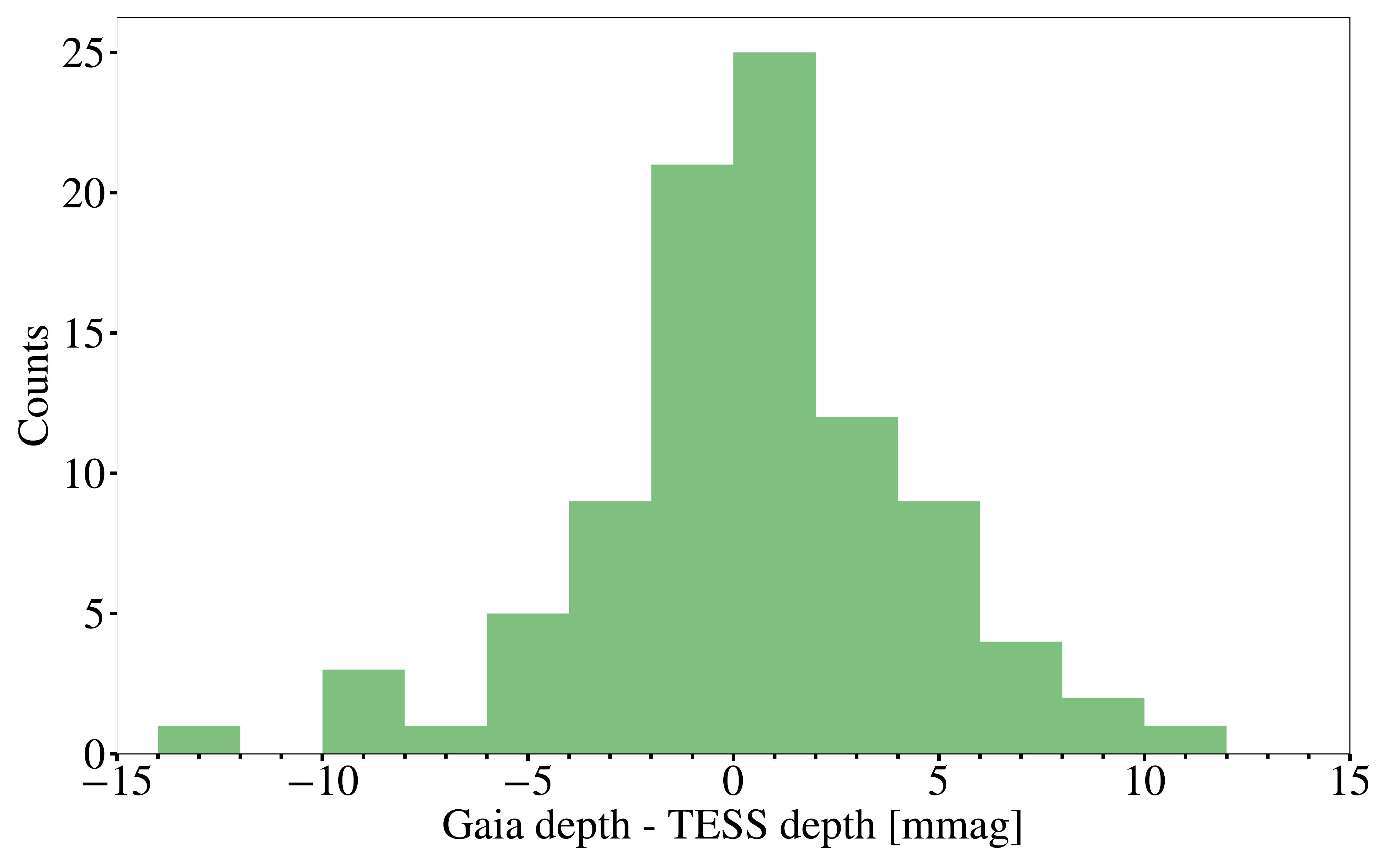}
\caption{Phase~II: Distribution of the differences between the Gaia and TESS transit depths 
for the $95$ on-target confirmations in Phase~II. The mean and standard deviation of these differences are $(\mu,\sigma) = (1.3,6.5)\,\mathrm{mmag}$.
}
\label{fig:Depths_differences} 
\end{figure}

%
Figure~\ref{fig:Depths_differences} presents a histogram of the differences between the transit depths as originally reported by TESS and the depths derived from the Gaia photometry, for the $95$ on-target confirmations found in Phase~II. The scatter is $\sigma_\mathrm{d} = 6.5$ mmag, and therefore we adopt this value  as the Gaia transit depth uncertainty for the on-target confirmations (Tables~\ref{tab:rest_params_conf} and  \ref{tab:group_params_conf}).

\begin{figure}
\centering
        \includegraphics[width=1.\linewidth]{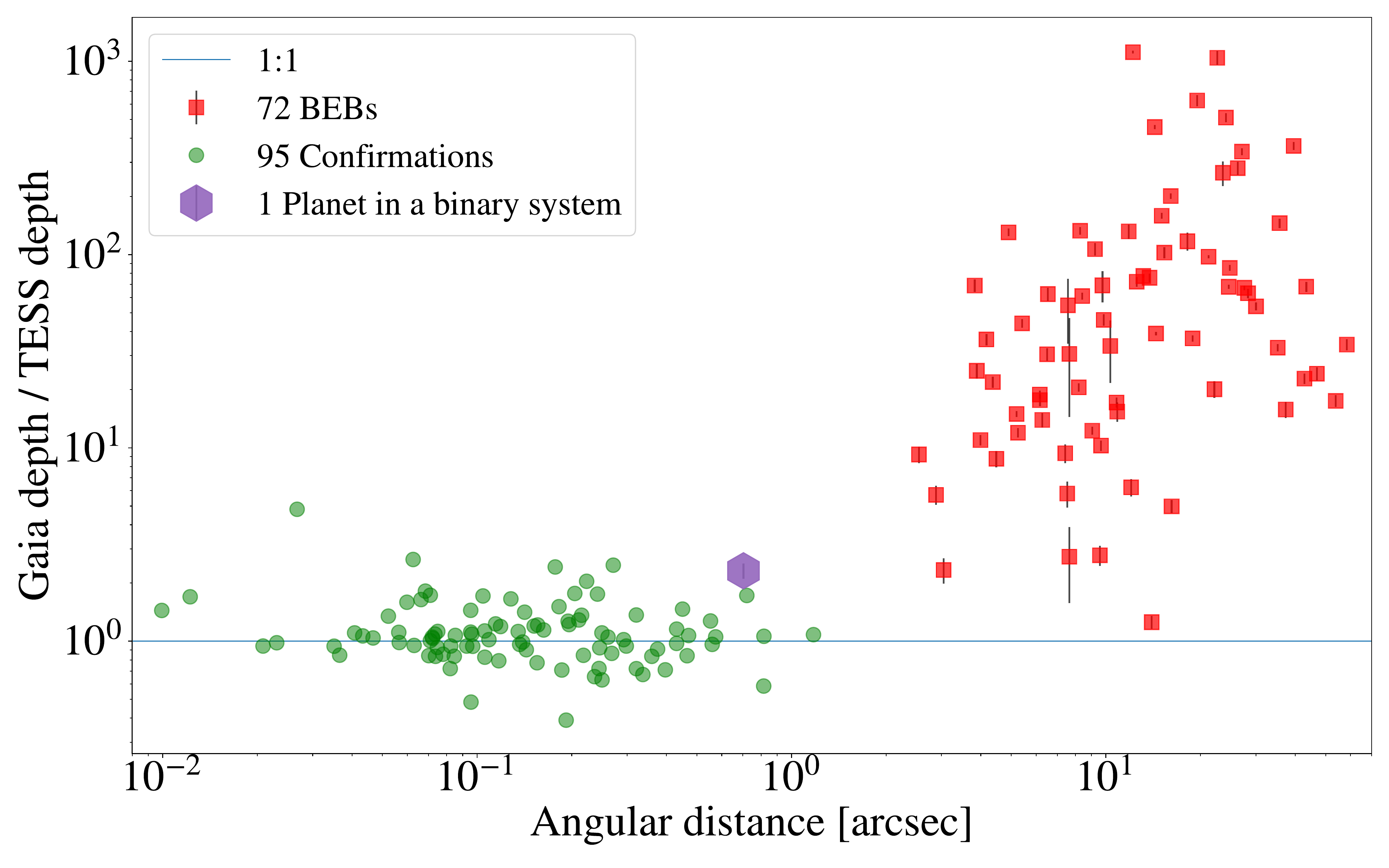}
        \caption{Phase~II: Transit depth ratios versus angular distance between the TOI targets and  the  Gaia detections, either BEBs or on-target confirmations. We omitted the transit-depth uncertainties for the on-target confirmations. 
        For most of the BEBs, the uncertainties are smaller than the square sizes.}
        \label{fig:General_DepthsRatiosAngDist} 
\end{figure}

Figure~\ref{fig:General_DepthsRatiosAngDist} presents the dependence of the ratio between the  Gaia and TESS depths on the angular distances between the identified Gaia sources and the corresponding TOI target stars. 
Naturally,  the depth seen by Gaia is substantially larger for BEBs.
As expected, the angular distance is small for on-target confirmations and large for BEBs, further delineating the separation between the two sets. 
Three out of the four BEB cases that had the smallest depth ratios were marked by the SG1 team as potential nearby planetary candidates (TOI-4211, TOI-4236 and TOI-4292). The BEB with the smallest depth ratio (TOI-3580) remained a BEB, as the stellar radius of the nearby host star is large enough to suggest a stellar eclipsing companion.

\section{Discussion \label{sec:discussion}}

This is an interim report on a collaborative effort of the {TESS} and {Gaia} teams to identify TOIs whose observed transit-like modulations are caused by faint neighboring BEBs and to confirm the TOI candidacy in some cases. The collaboration uses the growing database of Gaia epoch photometry and takes advantage of its high angular resolution and photometric precision, and has so far been able to confirm $\sim5\%$ of the TESS candidates as on-target and dismiss another $\sim5\%$ as BEBs. 

As the BEB modulations are large and their eclipses are deep, the eclipses are easily seen in the {Gaia} photometry, as demonstrated by Figs.~\ref{fig:BEB_4216} and \ref{fig:BEB_4346}. In a few cases where the eclipse was found to occur in a neighboring star, the depth of the undiluted eclipse seen in the Gaia data was shallow enough to still be caused by a planet orbiting the nearby star. In these cases the candidacy was shifted from the target star of the TOI.

The main limiting factor for identifying BEBs is the number of {Gaia} photometric measurements during transits, which was too small for most BEBs to allow a safe identification. 
%
%
In the sample of $1600$ TOIs, one could expect to have about 250 BEBs, while the search identified only $72$, simply because of the small number of {Gaia} photometric measurements. 

\begin{figure}
\centering
        \includegraphics[width=1\linewidth]{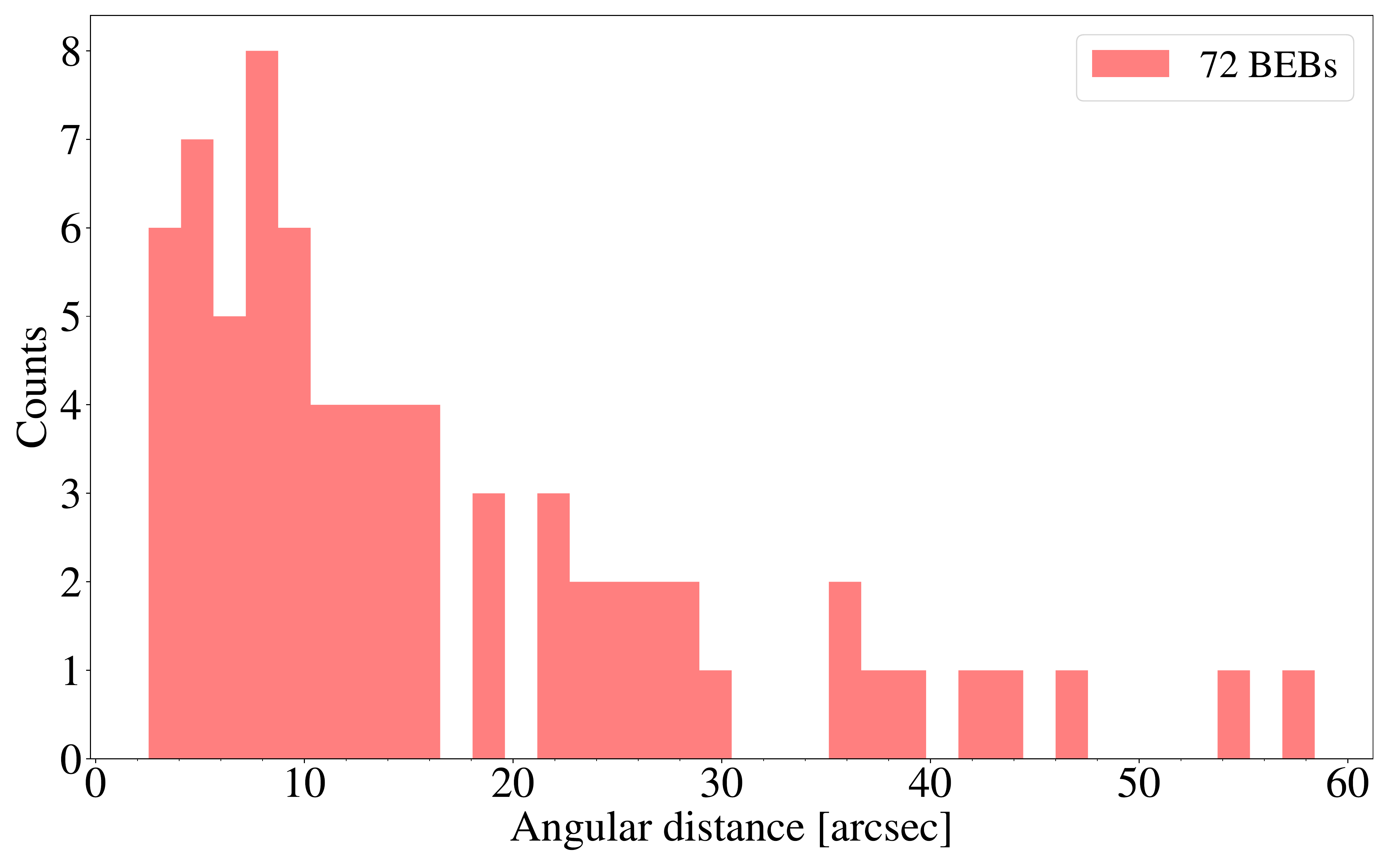}
        \caption{Phase~II: Angular distances of BEBs from their TOIs.}
        \label{fig:BEBs_AngDist_Histogram} 
\end{figure}

Figure~\ref{fig:BEBs_AngDist_Histogram} displays the angular distances of the detected BEBs in Phase~II. One can see that most of them were found within $40\as$ of their TOIs. Only $5$ out of $72$ BEBs are found in the range of $40$--$60\as$, indicating that the contamination drops substantially beyond $40\as$, equivalent to an area of approximately ten TESS pixels. We note that this is probably in accordance with  \cite[][see their Fig.~14]{Sullivan15}, who suggested an optimal collecting area of about ten pixels for a  star of 10 mag.
Furthermore, SG1 members privately reported that no BEBs further than $1\am$ have been found so far, probably because these cases were mostly rejected early on by the TESS team via the centroid analysis method. However, the $60\as$ radius might be too restrictive for a thorough search,  and we will slightly enlarge this radius in the next Gaia--TESS phase.

\begin{figure}
\centering
        \includegraphics[width=1\linewidth]{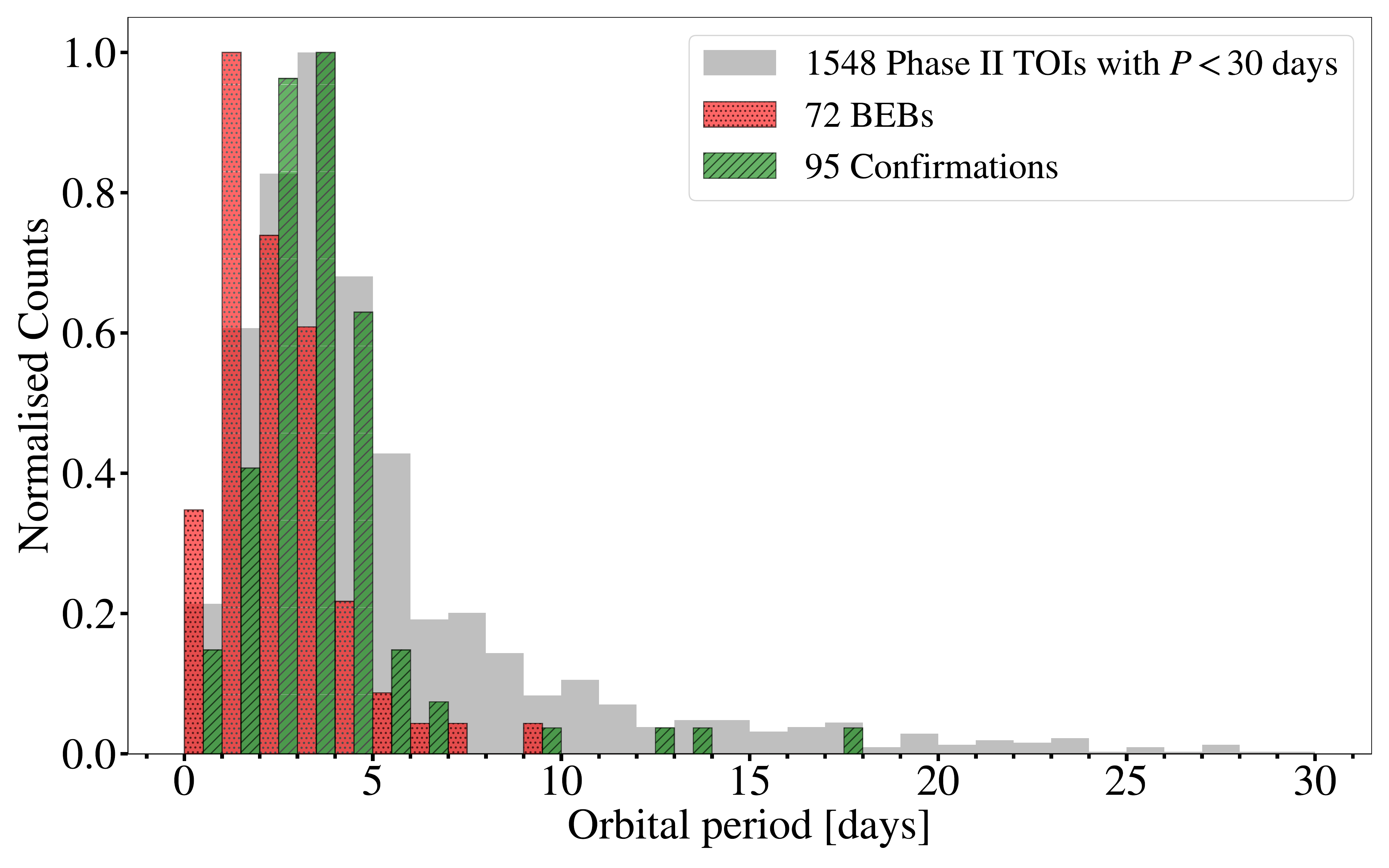}
        \caption{Phase~II: Normalized distribution of orbital periods for all TOIs up to a period of $30$ days (gray) and detected on-target confirmations (lined green) and BEBs (dotted red).}
        \label{fig:periods_Histogram} 
\end{figure}

Figure~\ref{fig:periods_Histogram} shows the distribution of orbital periods for our detections of on-target confirmations and BEBs relative to all TOIs in Phase~II, up to a period of $30$ days. Given the sparsity of the Gaia photometry, we expected to mostly detect cases with short orbital periods, as the fractional duration of a transit (duty cycle) declines with larger periods as $d/P\propto P^{-2/3}$ \citep{Sackett99}.

The Gaia data available to the collaboration consist of the first $34$ months of operation.
In the near future, the available photometry is expected to contain $66$ months of data, greatly increasing the capability and discovery rate of the present method, and possibly even allowing us to refine the ephemerides of some candidates.
When more {Gaia} measurements are available, we will be able to exclude more BEB alternatives. 
This will serve as a new way of confirming the on-target candidacy, allowing us to rule out all possible BEBs in neighboring stars and thus confirm the candidacy simply by elimination.

In the future, we plan to use the unpublished Gaia photometry, combined with the TESS photometry, to refine the TESS ephemerides, and to extend the BLS method to look for signs of transit-timing variations (TTV).

PLATO, the next transiting planet detection space mission, is expected to have a PSF of similar size to that of TESS \citep{Rauetal2016}. By the time PLATO is operational, Gaia photometry will have already become public, allowing the community to perform similar checks that will allow quick identification of genuine transit candidates and save valuable resources required for ground-based follow-up observations.

\begin{acknowledgements}
We deeply thank the referee, Alexandre Santerne, for reviewing this paper and for the enlightening and helpful comments that greatly helped us improve the manuscript.
This work has made use of data from the European Space Agency (ESA) mission Gaia (\url{https://www.cosmos.esa.int/gaia}), processed by the Gaia Data Processing and Analysis Consortium (DPAC, \url{https://www.cosmos.esa.int/web/gaia/dpac/consortium}). Funding for the DPAC has been provided by national institutions, some of which participate in the Gaia Multilateral Agreement.
T.M.\ acknowledges support of the United States-Israel Binational Science Foundation (BSF), grant No. 2016069. S.Z.\ acknowledges support by the Ministry of Science and Technology, Israel (grant 3-18143), and a grant from the Tel Aviv University Center for AI and Data Science (TAD). 
We have used the following \texttt{Python} code libraries: \texttt{Matplotlib} \citep{matplotlib}, \texttt{NumPy} \citep{numpy}, \texttt{AstroPy} \citep{astropy}, \texttt{Lightkurve} \citep{Lightkurve}, \texttt{TESScut.MAST} \citep{Astrocut}, \texttt{Eleanor} \citep{Eleanor} and \texttt{Pandas} \citep{Pandas_ADS}.
\end{acknowledgements}

\bibliographystyle{aa}
\bibliography{Bib.bib}

\begin{appendix}
\onecolumn
\section{Analysis results for Phase~I and Phase~II.}
\centering
\renewcommand{\arraystretch}{1.2}
{\tiny \tabcolsep=7pt 
\LTcapwidth=0.95\textwidth
\begin{longtable}[]{|lrrccc | rrrc|}
\caption{On-target confirmations in Phase~I search.$^{a}$}
\label{tab:rest_params_conf}\\
\hline
 \multicolumn{6}{| l |}{\textbf{TESS}} & \multicolumn{4}{| l |}{\textbf{Gaia}}\\
 \hline
  & \multicolumn{1}{ c }{TOI} & \multicolumn{1}{ c }{TIC} & Orbital Period & Transit Duration & Transit Depth &  \multicolumn{1}{ c }{$N$}  & \multicolumn{1}{ c }{$\smallsub{N}{IT}$} & $\mathcal{S/N}_\mathrm{T}$ & Transit Depth \\
    &  &     & [day]            & [day]              & [ppt]          &     &      &      &   [ppt] \\
 \hline
 \endfirsthead

 \hline
 \multicolumn{10}{|c|}{Table \ref{tab:rest_params_conf} continued.}\\
 \hline
 \multicolumn{6}{| l |}{\textbf{TESS}} & \multicolumn{4}{| c |}{\textbf{Gaia}}\\
 \hline
  & \multicolumn{1}{ c }{TOI} & \multicolumn{1}{ c }{TIC} & Orbital Period & Transit Duration & Transit Depth &  \multicolumn{1}{ c }{$N$}  & \multicolumn{1}{ c }{$\smallsub{N}{IT}$} & $\mathcal{S/N}_\mathrm{T}$ & Transit Depth \\
    &  &     & [day]            & [day]              & [ppt]          &     &      &      &   [ppt] \\
 \hline
 \endhead

 \hline
 \endfoot

 \hline
 \endlastfoot
1       & 2378.01       & 231063955     &  $ 3.49728 \pm 0.00037 $                       &  $ 0.0903 \pm 0.0045 $                        &  $ 17.35 \pm 0.63 $                              & 64    & 3     & 9.5                   & 21.1\\ 
2       & 2487.01       & 394346745     &  $ 3.40249 \pm 0.00010 $                       &  $ 0.1373 \pm 0.0026 $                        &  $ 14.15 \pm 0.22 $                              & 40    & 4     & 17.1                  & 12.9\\ 
3       & 2580.01       & 102713734     &  $ 3.39786 \pm 0.00019 $                       &  $ 0.1876 \pm 0.0052 $                        &  $ 9.96000 \pm 0.00093 $                           & 82    & 6     & 9.3                   & 9.0\\ 
4       & 2724.01       & 119290630     &  $ 1.48750550 \pm 5.0\cdot 10^{-7} $               &  $ 0.07808 \pm 0.00071 $                      &  $ 44.40000 \pm 0.00056 $                           & 40    & 3     & 13.1                  & 36.5\\ 
5       & 2757.01       & 147125706     &  $ 3.4813976 \pm 3.5\cdot 10^{-6} $               &  $ 0.1122 \pm 0.0045 $                        &  $ 21.8000 \pm 0.0054 $                            & 35    & 2     & 10.2                  & 21.2\\ 
6       & 2799.01       & 156887426     &  $ 1.43679140 \pm 9.0\cdot 10^{-7} $               &  $ 0.0908 \pm 0.0029 $                        &  $ 23.1800 \pm 0.0025 $                            & 39    & 5     & 14.0                  & 22.2\\ 
7       & 2818.01       & 151483286     &  $ 4.0397189 \pm 3.1\cdot 10^{-6} $               &  $ 0.1523 \pm 0.0025 $                        &  $ 15.11000 \pm 0.00090 $                           & 55    & 2     & 7.6                   & 12.7\\ 
8       & 2847.01       & 175604949     &  $ 2.6488450 \pm 2.1\cdot 10^{-6} $               &  $ 0.0888 \pm 0.0084 $                        &  $ 16.7700 \pm 0.0071 $                            & 44    & 2     & 7.9                   & 12.8\\ 
9       & 2916.01       & 76663549      &  $ 3.9216762 \pm 6.4\cdot 10^{-6} $               &  $ 0.0865 \pm 0.0034 $                        &  $ 17.4000 \pm 0.0069 $                            & 45    & 4     & 7.7                   & 22.5\\ 
10      & 3073.01       & 448744179     &  $ 3.14199 \pm 0.00053 $                       &  $ 0.082 \pm 0.013 $                          &  $ 13.310 \pm 0.018 $                             & 58    & 3     & 13.2                  & 10.5\\ 
11      & 3172.01       & 398502253     &  $ 3.00293 \pm 0.00096 $                       &  $ 0.119 \pm 0.023 $                          &  $ 5.080 \pm 0.022 $                             & 44    & 4     & 10.0                  & 8.7\\ 
12      & 3183.01       & 370385007     &  $ 4.2162 \pm 0.0020 $                         &  $ 0.160 \pm 0.016 $                          &  $ 6.330 \pm 0.013 $                             & 46    & 2     & 9.4                   & 13.9\\ 
13      & 3237.01       & 242389810     &  $ 3.02703 \pm 0.00045 $                       &  $ 0.1581 \pm 0.0077 $                        &  $ 21.5800 \pm 0.0088 $                            & 26    & 4     & 14.4                  & 20.6\\ 
14      & 3264.01       & 468055102     &  $ 3.313349 \pm 2.0\cdot 10^{-5} $               &  $ 0.1617 \pm 0.0062 $                        &  $ 8.9000 \pm 0.0026 $                            & 21    & 2     & 10.2                  & 14.1\\ 
15      & 3274.01       & 360822008     &  $ 2.907508 \pm 2.3\cdot 10^{-5} $               &  $ 0.1217 \pm 0.0093 $                        &  $ 10.0200 \pm 0.0087 $                            & 33    & 4     & 16.7                  & 12.7\\ 
16      & 3295.01       & 456574511     &  $ 3.4380 \pm 0.0010 $                         &  $ 0.161 \pm 0.013 $                          &  $ 8.370 \pm 0.010 $                             & 42    & 6     & 9.6                   & 11.8\\ 
17      & 3297.01       & 451538240     &  $ 3.93479 \pm 0.00033 $                       &  $ 0.185 \pm 0.014 $                          &  $ 8.1800 \pm 0.0044 $                            & 39    & 3     & 8.8                   & 7.7\\ 
18      & 3404.01       & 18795429      &  $ 2.5087868 \pm 4.2\cdot 10^{-6} $               &  $ 0.1054 \pm 0.0048 $                        &  $ 15.7500 \pm 0.0046 $                            & 68    & 3     & 13.4                  & 16.2\\ 
19      & 3406.01       & 410159011     &  $ 2.6233540 \pm 7.9\cdot 10^{-6} $               &  $ 0.119 \pm 0.014 $                          &  $ 8.7900 \pm 0.0053 $                            & 40    & 3     & 7.0                   & 8.2\\ 
20      & 3416.01       & 145910571     &  $ 2.3626917 \pm 8.2\cdot 10^{-6} $               &  $ 0.1224 \pm 0.0070 $                        &  $ 7.2800 \pm 0.0044 $                            & 54    & 5     & 8.0                   & 12.1\\ 
21      & 3441.01       & 393196631     &  $ 4.25899 \pm 0.00051 $                       &  $ 0.142 \pm 0.014 $                          &  $ 5.8900 \pm 0.0067 $                            & 48    & 3     & 9.0                   & 7.4\\ 
22      & 3444.01       & 161299119     &  $ 2.71943 \pm 0.00039 $                       &  $ 0.095 \pm 0.021 $                          &  $ 13.570 \pm 0.022 $                             & 67    & 2     & 7.0                   & 14.5\\ 
23      & 3458.01       & 340228388     &  $ 3.17305 \pm 0.00011 $                       &  $ 0.090 \pm 0.015 $                          &  $ 17.510 \pm 0.027 $                             & 50    & 2     & 13.6                  & 9.4\\ 
24      & 3473.01       & 406861922     &  $ 4.03676 \pm 0.00086 $                       &  $ 0.120 \pm 0.014 $                          &  $ 11.250 \pm 0.013 $                             & 44    & 3     & 14.4                  & 16.2\\ 
25      & 3474.01       & 274367763     &  $ 3.87893 \pm 0.00071 $                       &  $ 0.156 \pm 0.011 $                          &  $ 12.1200 \pm 0.0063 $                            & 37    & 3     & 8.4                   & 13.4\\ 
26      & 3477.01       & 402391147     &  $ 3.3614 \pm 0.0023 $                         &  $ 0.157 \pm 0.024 $                          &  $ 10.780 \pm 0.047 $                             & 72    & 3     & 16.5                  & 14.7\\ 
27      & 3479.01       & 402219289     &  $ 2.1968 \pm 0.0012 $                         &  $ 0.127 \pm 0.016 $                          &  $ 6.040 \pm 0.017 $                             & 46    & 4     & 8.8                   & 6.9\\ 
28      & 3484.01       & 221882319     &  $ 5.1905 \pm 0.0012 $                         &  $ 0.216 \pm 0.013 $                          &  $ 7.5700 \pm 0.0031 $                            & 38    & 2     & 7.8                   & 13.3\\ 
29      & 3491.01       & 208719443     &  $ 2.40389 \pm 0.00022 $                       &  $ 0.1515 \pm 0.0061 $                        &  $ 31.360 \pm 0.012 $                             & 42    & 5     & 20.7                  & 41.1\\ 
\hline
\multicolumn{10}{| l |}{\textbf{Transit confirmations in binary systems}}\\
\hline 
1       & 3277.01       & 319568619     &  $ 1.5265889 \pm 1.8\cdot 10^{-6} $               &  $ 0.0703 \pm 0.0025 $                        &  $ 23.6300 \pm 0.0031 $                            & 43    & 3     & 11.2                  & 29.8   
\end{longtable}
\begin{flushleft} 
\qquad$^{a}$ As described in Sect.~\ref{sec:transit_depths}, we adopt a uniform value of $\sigma_\mathrm{d} = 6.5$ mmag for the Gaia transit depth uncertainty of on-target confirmations.\\
\end{flushleft}
}


 {\tiny \tabcolsep=7pt 
\LTcapwidth=0.95\textwidth
\begin{longtable}[]{|lrrccc | rrrc|}
\caption{BEBs in Phase~I search.} 
\label{tab:rest_params_bebs}\\
 \hline
 \multicolumn{6}{| l |}{\textbf{TESS}} & \multicolumn{4}{| l |}{\textbf{Gaia}}\\
 \hline
  & \multicolumn{1}{ c }{TOI} & \multicolumn{1}{ c }{TIC} & Orbital Period & Transit Duration & Transit Depth &  \multicolumn{1}{ c }{$N$}  & \multicolumn{1}{ c }{$\smallsub{N}{IT}$} & $\mathcal{S/N}_\mathrm{T}$ & Transit Depth \\
    &  &     & [day]            & [day]              & [ppt]          &     &      &      &   [ppt] \\
 \hline
 \endfirsthead

 \hline
 \multicolumn{10}{|c|}{Table \ref{tab:rest_params_bebs} continued.}\\
 \hline
 \multicolumn{6}{| l |}{\textbf{TESS}} & \multicolumn{4}{| l |}{\textbf{Gaia}}\\
 \hline
  & \multicolumn{1}{ c }{TOI} & \multicolumn{1}{ c }{TIC} & Orbital Period & Transit Duration & Transit Depth &  \multicolumn{1}{ c }{$N$}  & \multicolumn{1}{ c }{$\smallsub{N}{IT}$} & $\mathcal{S/N}_\mathrm{T}$ & Transit Depth \\
    &  &     & [day]            & [day]              & [ppt]          &     &      &      &   [ppt] \\
 \hline
 \endhead

 \hline
 \endfoot

 \hline
 \endlastfoot
1       & 607.01        & 153651591     &  $ 2.14563 \pm 0.00030 $                       &  $ 0.090 \pm 0.013 $                          &  $ 1.44 \pm 0.10 $                              & 67    & 5     & 56.8                  &  $ 82.3 \pm 1.6 $                       \\ 
2       & 644.01        & 63303499      &  $ 1.92713 \pm 0.00021 $                       &  $ 0.190 \pm 0.016 $                          &  $ 0.586 \pm 0.035 $                             & 60    & 7     & 27.9                  &  $ 278 \pm 12 $                         \\ 
3       & 1323.01       & 256886630     &  $ 2.0390 \pm 0.0012 $                         &  $ 0.094 \pm 0.040 $                          &  $ 0.430 \pm 0.011 $                             & 72    & 2     & 24.9                  &  $ 569 \pm 34 $                         \\ 
4       & 1354.01       & 365683032     &  $ 1.42904 \pm 0.00048 $                       &  $ 0.137 \pm 0.019 $                          &  $ 1.21 \pm 0.16 $                              & 41    & 8     & 28.0                  &  $ 109.9 \pm 4.6 $                      \\ 
5       & 1374.01       & 190430205     &  $ 1.06884 \pm 0.00034 $                       &  $ 0.105 \pm 0.020 $                          &  $ 0.4700 \pm 0.0017 $                            & 52    & 6     & 23.7                  &  $ 177.3 \pm 8.7 $                      \\ 
6       & 1380.01       & 274170255     &  $ 1.41072 \pm 0.00028 $                       &  $ 0.109 \pm 0.017 $                          &  $ 0.5000 \pm 0.0011 $                            & 46    & 6     & 29.4                  &  $ 149.5 \pm 5.9 $                      \\ 
7       & 1395.01       & 468472950     &  $ 0.98736 \pm 0.00017 $                       &  $ 0.061 \pm 0.010 $                          &  $ 0.3700 \pm 0.0012 $                            & 51    & 2     & 14.4                  &  $ 83.8 \pm 6.2 $                       \\ 
8       & 1503.01       & 428065977     &  $ 1.65445 \pm 0.00067 $                       &  $ 0.133 \pm 0.028 $                          &  $ 0.6100 \pm 0.0020 $                            & 51    & 3     & 22.3                  &  $ 309 \pm 17 $                         \\ 
9       & 1510.01       & 377739433     &  $ 1.23852 \pm 0.00023 $                       &  $ 0.131 \pm 0.016 $                          &  $ 0.44000 \pm 0.00095 $                           & 41    & 4     & 22.0                  &  $ 334 \pm 19 $                         \\ 
10      & 1514.01       & 270238522     &  $ 1.36991 \pm 0.00028 $                       &  $ 0.173 \pm 0.012 $                          &  $ 1.1700 \pm 0.0012 $                            & 47    & 15    & 38.8                  &  $ 84.3 \pm 2.8 $                       \\ 
11      & 1557.01       & 320506985     &  $ 1.08697 \pm 0.00021 $                       &  $ 0.082 \pm 0.015 $                          &  $ 0.8600 \pm 0.0031 $                            & 76    & 5     & 9.2                   &  $ 38.6 \pm 4.4 $                       \\ 
12      & 1583.01       & 403256331     &  $ 3.77188 \pm 0.00076 $                       &  $ 0.122 \pm 0.019 $                          &  $ 0.4500 \pm 0.0010 $                            & 59    & 3     & 59.5                  &  $ 141.4 \pm 2.6 $                      \\ 
13      & 1950.01       & 327663613     &  $ 1.41144 \pm 0.00052 $                       &  $ 0.105 \pm 0.021 $                          &  $ 0.4144 \pm 0.0013 $                            & 47    & 3     & 36.4                  &  $ 363 \pm 13 $                         \\ 
14      & 1998.01       & 429295277     &  $ 1.10833 \pm 0.00032 $                       &  $ 0.095 \pm 0.014 $                          &  $ 1.0600 \pm 0.0018 $                            & 45    & 4     & 35.4                  &  $ 471 \pm 19 $                         \\ 
15      & 2004.01       & 439610306     &  $ 1.79988 \pm 0.00056 $                       &  $ 0.147 \pm 0.023 $                          &  $ 0.18000 \pm 0.00045 $                           & 52    & 2     & 11.6                  &  $ 79.0 \pm 7.2 $                       \\ 
16      & 2115.01       & 54464870      &  $ 3.69443 \pm 0.00019 $                       &  $ 0.194 \pm 0.023 $                          &  $ 0.21000 \pm 0.00040 $                           & 66    & 5     & 28.5                  &  $ 70.9 \pm 2.7 $                       \\ 
17      & 2517.01       & 48178229      &  $ 2.678825 \pm 2.8\cdot 10^{-5} $               &  $ 0.104 \pm 0.016 $                          &  $ 0.35000 \pm 0.00079 $                           & 72    & 6     & 39.4                  &  $ 194.1 \pm 5.7 $                      \\ 
18      & 2732.01       & 34466256      &  $ 0.7029205 \pm 1.9\cdot 10^{-6} $               &  $ 0.0552 \pm 0.0085 $                        &  $ 2.4800 \pm 0.0034 $                            & 35    & 2     & 17.3                  &  $ 384 \pm 28 $                         \\ 
19      & 2754.01       & 148233571     &  $ 0.9422588 \pm 6.0\cdot 10^{-6} $               &  $ 0.061 \pm 0.011 $                          &  $ 1.0600 \pm 0.0036 $                            & 34    & 2     & 14.0                  &  $ 195 \pm 16 $                         \\ 
20      & 2759.01       & 134315574     &  $ 1.5971584 \pm 9.2\cdot 10^{-6} $               &  $ 0.143 \pm 0.017 $                          &  $ 5.07 \pm 0.30 $                              & 31    & 2     & 30.1                  &  $ 173.5 \pm 6.5 $                      \\ 
21      & 2767.01       & 48806546      &  $ 0.9340570 \pm 5.2\cdot 10^{-6} $               &  $ 0.055 \pm 0.014 $                          &  $ 0.6200 \pm 0.0026 $                            & 84    & 3     & 32.9                  &  $ 388 \pm 15 $                         \\ 
22      & 2789.01       & 251209368     &  $ 3.007150 \pm 3.9\cdot 10^{-5} $               &  $ 0.186 \pm 0.023 $                          &  $ 3.3100 \pm 0.0079 $                            & 36    & 2     & 10.4                  &  $ 27.1 \pm 2.7 $                       \\ 
23      & 2805.01       & 5024743       &  $ 12.45086 \pm 0.00025 $                      &  $ 0.239 \pm 0.028 $                          &  $ 1.540 \pm 0.029 $                             & 58    & 2     & 23.8                  &  $ 94.9 \pm 4.3 $                       \\ 
24      & 2811.01       & 78775584      &  $ 5.555378 \pm 3.5\cdot 10^{-5} $               &  $ 0.094 \pm 0.014 $                          &  $ 1.0500 \pm 0.0025 $                            & 48    & 3     & 50.0                  &  $ 230.5 \pm 5.4 $                      \\ 
25      & 2812.01       & 105801221     &  $ 3.075073 \pm 3.3\cdot 10^{-5} $               &  $ 0.092 \pm 0.017 $                          &  $ 0.5100 \pm 0.0019 $                            & 61    & 2     & 45.3                  &  $ 333.8 \pm 9.1 $                      \\ 
26      & 2829.01       & 114103134     &  $ 1.1469027 \pm 2.7\cdot 10^{-6} $               &  $ 0.0643 \pm 0.0049 $                        &  $ 3.7 \pm 1.4 $                               & 49    & 2     & 29.3                  &  $ 175.4 \pm 6.7 $                      \\ 
27      & 2838.01       & 81231810      &  $ 5.452107 \pm 3.4\cdot 10^{-5} $               &  $ 0.196 \pm 0.031 $                          &  $ 8.170 \pm 0.011 $                             & 52    & 3     & 8.9                   &  $ 74.4 \pm 8.9 $                       \\ 
28      & 2878.01       & 4999813       &  $ 2.084952 \pm 2.2\cdot 10^{-5} $               &  $ 0.129 \pm 0.031 $                          &  $ 2.380 \pm 0.011 $                             & 25    & 2     & 14.6                  &  $ 42.1 \pm 3.1 $                       \\ 
29      & 2936.01       & 22020459      &  $ 1.332804 \pm 1.0\cdot 10^{-5} $               &  $ 0.115 \pm 0.012 $                          &  $ 1.9600 \pm 0.0026 $                            & 48    & 4     & 10.0                  &  $ 54.1 \pm 5.8 $                       \\ 
30      & 2998.01       & 34737413      &  $ 1.529712 \pm 1.5\cdot 10^{-5} $               &  $ 0.096 \pm 0.018 $                          &  $ 2.4100 \pm 0.0061 $                            & 49    & 5     & 96.2                  &  $ 278.4 \pm 3.6 $                      \\ 
31      & 3008.01       & 383666033     &  $ 3.967814 \pm 4.3\cdot 10^{-5} $               &  $ 0.177 \pm 0.025 $                          &  $ 0.7600 \pm 0.0016 $                            & 59    & 2     & 10.6                  &  $ 39.1 \pm 3.8 $                       \\ 
32      & 3018.01       & 45684223      &  $ 6.583202 \pm 3.0\cdot 10^{-5} $               &  $ 0.131 \pm 0.011 $                          &  $ 5.2800 \pm 0.0040 $                            & 53    & 2     & 28.2                  &  $ 133.8 \pm 5.2 $                      \\ 
33      & 3064.01       & 463070290     &  $ 0.6771705 \pm 2.0\cdot 10^{-6} $               &  $ 0.0594 \pm 0.0096 $                        &  $ 1.1400 \pm 0.0015 $                            & 36    & 2     & 16.7                  &  $ 118.0 \pm 7.7 $                      \\ 
34      & 3258.01       & 369789627     &  $ 1.22785 \pm 0.00010 $                       &  $ 0.071 \pm 0.014 $                          &  $ 2.0700 \pm 0.0056 $                            & 38    & 4     & 12.2                  &  $ 189 \pm 18 $                         \\ 
35      & 3347.01       & 131581531     &  $ 3.059164 \pm 2.8\cdot 10^{-5} $               &  $ 0.177 \pm 0.025 $                          &  $ 2.1100 \pm 0.0035 $                            & 52    & 3     & 12.6                  &  $ 51.5 \pm 4.3 $                       \\ 
36      & 3357.01       & 412259468     &  $ 2.65358 \pm 0.00038 $                       &  $ 0.088 \pm 0.015 $                          &  $ 1.45 \pm 0.11 $                              & 52    & 2     & 15.7                  &  $ 141.0 \pm 9.8 $                      \\ 
37      & 3369.01       & 190981760     &  $ 3.546358 \pm 2.0\cdot 10^{-5} $               &  $ 0.174 \pm 0.026 $                          &  $ 6.0200 \pm 0.0084 $                            & 56    & 3     & 64.6                  &  $ 122.8 \pm 2.1 $                      \\ 
38      & 3375.01       & 185482513     &  $ 1.236832 \pm 1.0\cdot 10^{-5} $               &  $ 0.089 \pm 0.020 $                          &  $ 1.9300 \pm 0.0069 $                            & 55    & 2     & 13.1                  &  $ 43.4 \pm 3.4 $                       \\ 
39      & 3377.01       & 184670625     &  $ 0.5241284 \pm 2.1\cdot 10^{-6} $               &  $ 0.0424 \pm 0.0073 $                        &  $ 2.4 \pm 1.2 $                               & 43    & 2     & 23.1                  &  $ 267 \pm 14 $                         \\ 
40      & 3385.01       & 175079720     &  $ 2.262326 \pm 1.1\cdot 10^{-5} $               &  $ 0.127 \pm 0.025 $                          &  $ 2.9400 \pm 0.0060 $                            & 41    & 2     & 14.0                  &  $ 51.7 \pm 3.9 $                       \\ 
41      & 3388.01       & 173860688     &  $ 0.84566 \pm 0.00026 $                       &  $ 0.076 \pm 0.016 $                          &  $ 0.3900 \pm 0.0020 $                            & 36    & 3     & 28.7                  &  $ 380 \pm 17 $                         \\ 
42      & 3392.01       & 80174172      &  $ 1.1104301 \pm 6.0\cdot 10^{-6} $               &  $ 0.067 \pm 0.014 $                          &  $ 4.870 \pm 0.015 $                             & 38    & 4     & 23.4                  &  $ 282 \pm 15 $                         \\ 
43      & 3399.01       & 153197346     &  $ 2.64245 \pm 0.00071 $                       &  $ 0.182 \pm 0.022 $                          &  $ 4.700 \pm 0.011 $                             & 55    & 3     & 8.0                   &  $ 23.5 \pm 3.0 $                       \\ 
44      & 3402.01       & 152685831     &  $ 3.471417 \pm 2.8\cdot 10^{-5} $               &  $ 0.125 \pm 0.038 $                          &  $ 3.580 \pm 0.017 $                             & 46    & 3     & 16.6                  &  $ 140.5 \pm 9.4 $                      \\ 
45      & 3420.01       & 293978434     &  $ 1.59714 \pm 0.00042 $                       &  $ 0.138 \pm 0.019 $                          &  $ 2.1500 \pm 0.0045 $                            & 55    & 4     & 20.8                  &  $ 234 \pm 13 $                         \\ 
46      & 3423.01       & 144970275     &  $ 2.531342 \pm 1.0\cdot 10^{-5} $               &  $ 0.0936 \pm 0.0095 $                        &  $ 5.8300 \pm 0.0075 $                            & 50    & 4     & 31.6                  &  $ 219.5 \pm 8.1 $                      \\ 
47      & 3424.01       & 283702603     &  $ 0.6691882 \pm 2.0\cdot 10^{-6} $               &  $ 0.056 \pm 0.012 $                          &  $ 2.8500 \pm 0.0078 $                            & 74    & 5     & 26.1                  &  $ 266 \pm 12 $                         \\ 
48      & 3436.01       & 132320310     &  $ 1.3955317 \pm 7.7\cdot 10^{-6} $               &  $ 0.109 \pm 0.016 $                          &  $ 0.9600 \pm 0.0018 $                            & 37    & 2     & 16.4                  &  $ 219 \pm 15 $                         \\ 
49      & 3440.01       & 208030798     &  $ 1.7168 \pm 0.0013 $                         &  $ 0.085 \pm 0.015 $                          &  $ 13.620 \pm 0.055 $                             & 59    & 3     & 22.3                  &  $ 475 \pm 29 $                         \\ 
50      & 3466.01       & 284254253     &  $ 1.8629 \pm 0.0020 $                         &  $ 0.190 \pm 0.016 $                          &  $ 3.4 \pm 1.4 $                               & 61    & 8     & 22.4                  &  $ 49.1 \pm 2.4 $                       \\ 
51      & 3467.01       & 453289469     &  $ 0.779919 \pm 5.3\cdot 10^{-5} $               &  $ 0.0643 \pm 0.0082 $                        &  $ 1.9200 \pm 0.0035 $                            & 40    & 2     & 9.2                   &  $ 182 \pm 22 $                         \\ 
52      & 3472.01       & 275562670     &  $ 1.58106 \pm 0.00035 $                       &  $ 0.1015 \pm 0.0082 $                        &  $ 33.340 \pm 0.038 $                             & 44    & 2     & 21.9                  &  $ 197 \pm 10 $                         \\ 
 \end{longtable}}
\twocolumn

\onecolumn
\centering
\renewcommand{\arraystretch}{1.2}
{\tiny \tabcolsep=7pt 
\LTcapwidth=0.95\textwidth
\begin{longtable}[]{|lrrccc | rrrc|}
\caption{On-target confirmations in Phase~II.$^{b}$}
\label{tab:group_params_conf}\\
\hline
 \multicolumn{6}{| l |}{\textbf{TESS}} & \multicolumn{4}{| l |}{\textbf{Gaia}}\\
 \hline
  & \multicolumn{1}{ c }{TOI} & \multicolumn{1}{ c }{TIC} & Orbital Period & Transit Duration & Transit Depth &  \multicolumn{1}{ c }{$N$}  & \multicolumn{1}{ c }{$\smallsub{N}{IT}$} & $\mathcal{S/N}_\mathrm{T}$ & Transit Depth \\
    &  &     & [day]            & [day]              & [ppt]          &     &      &      &   [ppt] \\
 \hline
 \endfirsthead

 \hline
 \multicolumn{10}{|c|}{Table \ref{tab:group_params_conf} continued.}\\
 \hline
 \multicolumn{6}{| l |}{\textbf{TESS}} & \multicolumn{4}{| l |}{\textbf{Gaia}}\\
\hline
  & \multicolumn{1}{ c }{TOI} & \multicolumn{1}{ c }{TIC} & Orbital Period & Transit Duration & Transit Depth &  \multicolumn{1}{ c }{$N$}  & \multicolumn{1}{ c }{$\smallsub{N}{IT}$} & $\mathcal{S/N}_\mathrm{T}$ & Transit Depth \\
    &  &     & [day]            & [day]              & [ppt]          &     &      &      &   [ppt] \\
 \hline
 \endhead

 \hline
 \endfoot

 \hline
 \endlastfoot

1       & 3510.01       & 57753734      &  $ 2.8729 \pm 0.0011 $                         &  $ 0.111 \pm 0.014 $                          &  $ 11.760 \pm 0.023 $                             & 50    & 3     & 17.4                  & 14.2\\ 
2       & 3528.01       & 67247610      &  $ 1.97480 \pm 0.00037 $                       &  $ 0.065 \pm 0.016 $                          &  $ 11.780 \pm 0.054 $                             & 42    & 4     & 8.5                   & 20.2\\ 
3       & 3531.01       & 316038054     &  $ 3.80974 \pm 0.00029 $                       &  $ 0.181 \pm 0.015 $                          &  $ 13.4800 \pm 0.0034 $                            & 50    & 7     & 18.6                  & 12.3\\ 
4       & 3558.01       & 195750008     &  $ 2.5678 \pm 0.0014 $                         &  $ 0.085 \pm 0.016 $                          &  $ 12.080 \pm 0.042 $                             & 51    & 3     & 9.3                   & 9.4\\ 
5       & 3582.01       & 290308758     &  $ 4.41685 \pm 0.00026 $                       &  $ 0.160 \pm 0.011 $                          &  $ 10.6200 \pm 0.0076 $                            & 43    & 2     & 9.2                   & 10.2\\ 
6       & 3599.01       & 194461202     &  $ 3.25231 \pm 0.00066 $                       &  $ 0.113 \pm 0.011 $                          &  $ 11.410 \pm 0.011 $                             & 70    & 2     & 7.3                   & 8.1\\ 
7       & 3602.01       & 331146317     &  $ 2.38761 \pm 0.00016 $                       &  $ 0.1098 \pm 0.0060 $                        &  $ 14.7000 \pm 0.0073 $                            & 52    & 2     & 13.7                  & 14.0\\ 
8       & 3622.01       & 432761635     &  $ 3.21300 \pm 0.00092 $                       &  $ 0.140 \pm 0.014 $                          &  $ 10.330 \pm 0.013 $                             & 46    & 4     & 9.9                   & 13.1\\ 
9       & 3635.01       & 428003305     &  $ 1.80427 \pm 0.00027 $                       &  $ 0.1722 \pm 0.0098 $                        &  $ 4.3000 \pm 0.0035 $                            & 64    & 4     & 9.5                   & 7.3\\ 
10      & 3646.01       & 445407434     &  $ 2.17959 \pm 0.00013 $                       &  $ 0.1089 \pm 0.0050 $                        &  $ 19.4600 \pm 0.0069 $                            & 84    & 3     & 15.3                  & 28.0\\ 
11      & 3664.01       & 348437470     &  $ 3.2963 \pm 0.0011 $                         &  $ 0.093 \pm 0.013 $                          &  $ 13.000 \pm 0.035 $                             & 39    & 2     & 8.2                   & 9.2\\ 
12      & 3673.01       & 452964680     &  $ 2.23910 \pm 0.00015 $                       &  $ 0.1224 \pm 0.0077 $                        &  $ 15.1200 \pm 0.0074 $                            & 76    & 3     & 12.5                  & 20.3\\ 
13      & 3685.01       & 252118701     &  $ 4.97848 \pm 0.00067 $                       &  $ 0.117 \pm 0.010 $                          &  $ 10.540 \pm 0.014 $                             & 81    & 2     & 8.2                   & 16.7\\ 
14      & 3699.01       & 266401846     &  $ 3.04416 \pm 0.00057 $                       &  $ 0.121 \pm 0.013 $                          &  $ 11.1800 \pm 0.0098 $                            & 45    & 4     & 15.1                  & 11.6\\ 
15      & 3703.01       & 260962960     &  $ 3.8881 \pm 0.0014 $                         &  $ 0.166 \pm 0.019 $                          &  $ 8.5600 \pm 0.0063 $                            & 39    & 2     & 6.9                   & 9.3\\ 
16      & 3706.01       & 252430813     &  $ 4.3721 \pm 0.0011 $                         &  $ 0.1631 \pm 0.0085 $                        &  $ 13.800 \pm 0.010 $                             & 53    & 3     & 10.4                  & 11.4\\ 
17      & 3714.01       & 155867025     &  $ 2.15483 \pm 0.00015 $                       &  $ 0.0645 \pm 0.0028 $                        &  $ 45.880 \pm 0.014 $                             & 59    & 2     & 9.7                   & 49.5\\ 
18      & 3719.01       & 150074121     &  $ 1.22203 \pm 0.00018 $                       &  $ 0.0685 \pm 0.0083 $                        &  $ 8.730 \pm 0.011 $                             & 73    & 5     & 7.7                   & 7.4\\ 
19      & 3749.01       & 284174392     &  $ 1.87541 \pm 0.00036 $                       &  $ 0.1241 \pm 0.0083 $                        &  $ 9.41 \pm 0.49 $                              & 30    & 2     & 8.6                   & 12.8\\ 
20      & 3751.01       & 284173938     &  $ 2.91544 \pm 0.00043 $                       &  $ 0.1479 \pm 0.0058 $                        &  $ 21.31 \pm 0.68 $                              & 29    & 3     & 10.2                  & 10.4\\ 
21      & 3758.01       & 280254984     &  $ 3.41769 \pm 0.00017 $                       &  $ 0.1443 \pm 0.0061 $                        &  $ 10.4700 \pm 0.0032 $                            & 53    & 4     & 7.9                   & 10.7\\ 
22      & 3780.01       & 468832904     &  $ 1.4005325 \pm 2.7\cdot 10^{-6} $               &  $ 0.0863 \pm 0.0030 $                        &  $ 21.9700 \pm 0.0098 $                            & 61    & 7     & 19.6                  & 23.4\\ 
23      & 3787.01       & 457104362     &  $ 4.81500 \pm 0.00081 $                       &  $ 0.153 \pm 0.011 $                          &  $ 15.4000 \pm 0.0080 $                            & 52    & 2     & 10.1                  & 11.1\\ 
24      & 3791.01       & 400432230     &  $ 3.1833866 \pm 2.5\cdot 10^{-6} $               &  $ 0.0841 \pm 0.0032 $                        &  $ 9.720 \pm 0.012 $                             & 43    & 2     & 10.3                  & 10.7\\ 
25      & 3814.01       & 155873992     &  $ 4.951310 \pm 1.4\cdot 10^{-5} $               &  $ 0.1622 \pm 0.0081 $                        &  $ 8.7800 \pm 0.0038 $                            & 67    & 4     & 6.7                   & 10.1\\ 
26      & 3842.01       & 165985431     &  $ 3.65662 \pm 0.00027 $                       &  $ 0.1065 \pm 0.0048 $                        &  $ 21.7300 \pm 0.0084 $                            & 52    & 3     & 16.8                  & 13.8\\ 
27      & 3901.01       & 357373216     &  $ 9.54685 \pm 0.00011 $                       &  $ 0.097 \pm 0.018 $                          &  $ 19.050 \pm 0.029 $                             & 48    & 2     & 16.8                  & 16.0\\ 
28      & 3915.01       & 81247877      &  $ 2.201376 \pm 9.4\cdot 10^{-5} $               &  $ 0.0932 \pm 0.0043 $                        &  $ 25.6300 \pm 0.0043 $                            & 64    & 4     & 8.9                   & 27.1\\ 
29      & 3946.01       & 336399144     &  $ 5.703341 \pm 3.8\cdot 10^{-5} $               &  $ 0.1266 \pm 0.0087 $                        &  $ 14.7300 \pm 0.0065 $                            & 41    & 3     & 9.2                   & 12.3\\ 
30      & 3947.01       & 336092760     &  $ 6.77151 \pm 0.00029 $                       &  $ 0.264 \pm 0.028 $                          &  $ 3.4000 \pm 0.0049 $                            & 38    & 5     & 8.9                   & 6.2\\ 
31      & 3952.01       & 321861405     &  $ 2.989774 \pm 1.4\cdot 10^{-5} $               &  $ 0.1347 \pm 0.0055 $                        &  $ 13.9000 \pm 0.0024 $                            & 42    & 2     & 8.8                   & 11.8\\ 
32      & 3955.01       & 314375831     &  $ 3.127759 \pm 2.0\cdot 10^{-5} $               &  $ 0.0936 \pm 0.0066 $                        &  $ 13.5100 \pm 0.0051 $                            & 49    & 3     & 7.2                   & 13.3\\ 
33      & 3965.01       & 269043098     &  $ 4.72540 \pm 0.00012 $                       &  $ 0.209 \pm 0.014 $                          &  $ 6.3200 \pm 0.0036 $                            & 44    & 3     & 10.1                  & 9.1\\ 
34      & 3974.01       & 156007004     &  $ 1.792224 \pm 5.6\cdot 10^{-5} $               &  $ 0.0701 \pm 0.0040 $                        &  $ 18.2700 \pm 0.0054 $                            & 46    & 3     & 8.7                   & 23.2\\ 
35      & 3978.01       & 468329664     &  $ 3.547237 \pm 9.3\cdot 10^{-5} $               &  $ 0.187 \pm 0.014 $                          &  $ 6.5800 \pm 0.0039 $                            & 41    & 4     & 8.9                   & 6.2\\ 
36      & 4000.01       & 283015592     &  $ 3.702766 \pm 5.2\cdot 10^{-5} $               &  $ 0.1483 \pm 0.0080 $                        &  $ 11.2200 \pm 0.0052 $                            & 50    & 2     & 8.3                   & 12.5\\ 
37      & 4003.01       & 399346877     &  $ 3.960696 \pm 8.0\cdot 10^{-5} $               &  $ 0.140 \pm 0.011 $                          &  $ 5.8800 \pm 0.0046 $                            & 64    & 2     & 9.1                   & 9.7\\ 
38      & 4007.01       & 389910188     &  $ 4.543204 \pm 4.8\cdot 10^{-5} $               &  $ 0.1555 \pm 0.0059 $                        &  $ 12.3100 \pm 0.0043 $                            & 64    & 3     & 6.5                   & 13.9\\ 
39      & 4012.01       & 280315875     &  $ 1.825438 \pm 1.1\cdot 10^{-5} $               &  $ 0.0738 \pm 0.0040 $                        &  $ 21.270 \pm 0.011 $                             & 77    & 4     & 7.9                   & 31.0\\ 
40      & 4034.01       & 375654303     &  $ 1.802096 \pm 1.0\cdot 10^{-5} $               &  $ 0.1254 \pm 0.0065 $                        &  $ 8.7400 \pm 0.0021 $                            & 41    & 7     & 10.3                  & 8.6\\ 
41      & 4069.01       & 143009290     &  $ 2.72342 \pm 0.00056 $                       &  $ 0.073 \pm 0.015 $                          &  $ 13.090 \pm 0.032 $                             & 49    & 4     & 14.0                  & 12.4\\ 
42      & 4077.01       & 57159675      &  $ 5.56878 \pm 0.00042 $                       &  $ 0.0849 \pm 0.0069 $                        &  $ 12.330 \pm 0.010 $                             & 51    & 3     & 8.7                   & 8.9\\ 
43      & 4080.01       & 428892437     &  $ 3.751998 \pm 1.3\cdot 10^{-5} $               &  $ 0.126 \pm 0.014 $                          &  $ 18.200 \pm 0.019 $                             & 44    & 3     & 13.3                  & 12.0\\ 
44      & 4104.01       & 235664497     &  $ 2.37636 \pm 0.00065 $                       &  $ 0.1411 \pm 0.0094 $                        &  $ 7.2700 \pm 0.0085 $                            & 74    & 4     & 5.6                   & 7.6\\ 
45      & 4115.01       & 199610140     &  $ 2.985597 \pm 1.5\cdot 10^{-5} $               &  $ 0.0517 \pm 0.0059 $                        &  $ 9.720 \pm 0.099 $                             & 37    & 2     & 9.0                   & 10.2\\ 
46      & 4147.01       & 138329479     &  $ 2.374539 \pm 2.2\cdot 10^{-5} $               &  $ 0.1351 \pm 0.0069 $                        &  $ 10.2600 \pm 0.0040 $                            & 51    & 4     & 8.8                   & 9.3\\ 
47      & 4160.01       & 441580526     &  $ 4.442746 \pm 2.6\cdot 10^{-5} $               &  $ 0.1353 \pm 0.0053 $                        &  $ 13.2900 \pm 0.0030 $                            & 61    & 3     & 8.0                   & 15.2\\ 
48      & 4181.01       & 426017070     &  $ 0.5992462 \pm 4.1\cdot 10^{-6} $               &  $ 0.0923 \pm 0.0046 $                        &  $ 23.3 \pm 1.2 $                               & 34    & 4     & 8.6                   & 27.9\\ 
49      & 4209.01       & 200090347     &  $ 0.86317 \pm 0.00020 $                       &  $ 0.0511 \pm 0.0092 $                        &  $ 6.330 \pm 0.025 $                             & 68    & 4     & 8.8                   & 6.0\\ 
50      & 4218.01       & 142628514     &  $ 2.62695 \pm 0.00045 $                       &  $ 0.1460 \pm 0.0063 $                        &  $ 12.0100 \pm 0.0045 $                            & 72    & 2     & 5.9                   & 13.1\\ 
51      & 4262.01       & 53365065      &  $ 5.997789 \pm 1.2\cdot 10^{-5} $               &  $ 0.1801 \pm 0.0066 $                        &  $ 7.7500 \pm 0.0010 $                            & 37    & 3     & 8.0                   & 7.5\\ 
52      & 4274.01       & 290388627     &  $ 0.445962 \pm 3.1\cdot 10^{-5} $               &  $ 0.0566 \pm 0.0083 $                        &  $ 11.500 \pm 0.024 $                             & 49    & 5     & 7.4                   & 9.1\\ 
53      & 4280.01       & 387193925     &  $ 3.518972 \pm 1.8\cdot 10^{-5} $               &  $ 0.1338 \pm 0.0086 $                        &  $ 13.8200 \pm 0.0063 $                            & 32    & 2     & 9.5                   & 13.0\\ 
54      & 4282.01       & 374348168     &  $ 13.509804 \pm 3.6\cdot 10^{-5} $               &  $ 0.1123 \pm 0.0070 $                        &  $ 24.500 \pm 0.032 $                             & 67    & 2     & 6.5                   & 16.5\\ 
55      & 4283.01       & 343952976     &  $ 3.013092 \pm 4.4\cdot 10^{-5} $               &  $ 0.0976 \pm 0.0072 $                        &  $ 14.1500 \pm 0.0036 $                            & 45    & 2     & 12.8                  & 14.8\\ 
56      & 4285.01       & 310105752     &  $ 1.170544 \pm 9.5\cdot 10^{-5} $               &  $ 0.057 \pm 0.010 $                          &  $ 4.890 \pm 0.013 $                             & 49    & 5     & 6.6                   & 4.6\\ 
57      & 4289.01       & 278540265     &  $ 2.9757432 \pm 9.6\cdot 10^{-6} $               &  $ 0.1864 \pm 0.0077 $                        &  $ 6.98000 \pm 0.00057 $                           & 38    & 5     & 7.8                   & 11.4\\ 
58      & 4293.01       & 101696403     &  $ 1.6242109 \pm 4.1\cdot 10^{-6} $               &  $ 0.0947 \pm 0.0025 $                        &  $ 17.7400 \pm 0.0019 $                            & 45    & 3     & 14.0                  & 19.0\\ 
59      & 4329.01       & 256722647     &  $ 2.922470 \pm 3.0\cdot 10^{-5} $               &  $ 0.2100 \pm 0.0087 $                        &  $ 4.18000 \pm 0.00081 $                           & 43    & 5     & 7.7                   & 7.2\\ 
60      & 4381.01       & 305767364     &  $ 1.497422 \pm 2.1\cdot 10^{-5} $               &  $ 0.09266 \pm 0.00066 $                      &  $ 39.32 \pm 0.31 $                              & 52    & 4     & 5.3                   & 34.1\\ 
61      & 4439.01       & 267545252     &  $ 3.37921 \pm 0.00042 $                       &  $ 0.1230 \pm 0.0063 $                        &  $ 18.14 \pm 0.66 $                              & 31    & 2     & 20.6                  & 17.5\\ 
62      & 4442.01       & 446158352     &  $ 4.04318 \pm 0.00084 $                       &  $ 0.1627 \pm 0.0094 $                        &  $ 10.97 \pm 0.56 $                              & 39    & 2     & 7.4                   & 9.4\\ 
63      & 4463.01       & 8599009           &  $ 2.8807180 \pm 4.3\cdot 10^{-6} $               &  $ 0.0647 \pm 0.0089 $                        &  $ 11.1600 \pm 0.0097 $                            & 40    & 3     & 8.5                   & 6.6\\ 
64      & 4468.01       & 441763252     &  $ 2.77090 \pm 0.00017 $                       &  $ 0.0959 \pm 0.0032 $                        &  $ 20.70 \pm 0.45 $                              & 35    & 3     & 19.4                  & 20.5\\ 
65      & 4497.01       & 436166416     &  $ 3.600219 \pm 1.2\cdot 10^{-5} $               &  $ 0.148 \pm 0.013 $                          &  $ 8.15 \pm 0.22 $                              & 72    & 2     & 5.6                   & 7.5\\ 
66      & 4666.01       & 165202476     &  $ 2.90939 \pm 0.00019 $                       &  $ 0.0865 \pm 0.0042 $                        &  $ 45.160 \pm 0.016 $                             & 50    & 2     & 4.6                   & 37.8\\ 
67      & 4674.01       & 89062142      &  $ 4.243063 \pm 2.2\cdot 10^{-5} $               &  $ 0.149 \pm 0.012 $                          &  $ 7.2700 \pm 0.0059 $                            & 60    & 3     & 2.9                   & 2.8\\ 
68      & 4678.01       & 206717400     &  $ 4.774317 \pm 2.4\cdot 10^{-5} $               &  $ 0.124 \pm 0.014 $                          &  $ 10.690 \pm 0.011 $                             & 27    & 2     & 6.6                   & 7.7\\ 
69      & 4680.01       & 31258738      &  $ 17.43140 \pm 0.00015 $                      &  $ 0.169 \pm 0.013 $                          &  $ 8.1800 \pm 0.0089 $                            & 55    & 2     & 6.4                   & 10.5\\ 
70      & 4720.01       & 124322274     &  $ 3.781083 \pm 2.4\cdot 10^{-5} $               &  $ 0.128 \pm 0.013 $                          &  $ 8.030 \pm 0.012 $                             & 27    & 2     & 4.8                   & 9.0\\ 
71      & 4754.01       & 61344769      &  $ 3.630728 \pm 1.2\cdot 10^{-5} $               &  $ 0.133 \pm 0.015 $                          &  $ 10.5500 \pm 0.0081 $                            & 33    & 4     & 14.3                  & 9.8\\ 
72      & 4761.01       & 440849210     &  $ 3.786305 \pm 4.6\cdot 10^{-5} $               &  $ 0.199 \pm 0.021 $                          &  $ 2.1400 \pm 0.0036 $                            & 23    & 2     & 4.5                   & 5.7\\ 
73      & 4777.01       & 170368805     &  $ 2.958876 \pm 1.1\cdot 10^{-5} $               &  $ 0.112 \pm 0.027 $                          &  $ 12.90 \pm 0.17 $                              & 56    & 5     & 13.6                  & 13.0\\ 
74      & 4778.01       & 145163386     &  $ 3.807421 \pm 2.2\cdot 10^{-5} $               &  $ 0.166 \pm 0.016 $                          &  $ 7.6700 \pm 0.0069 $                            & 43    & 8     & 13.3                  & 6.4\\ 
75      & 4793.01       & 65949719      &  $ 3.465863 \pm 2.3\cdot 10^{-5} $               &  $ 0.138 \pm 0.012 $                          &  $ 9.580 \pm 0.011 $                             & 45    & 4     & 5.9                   & 10.0\\ 
76      & 4797.01       & 265011923     &  $ 2.9922495 \pm 8.5\cdot 10^{-6} $               &  $ 0.0664 \pm 0.0076 $                        &  $ 10.895 \pm 0.018 $                             & 32    & 2     & 2.7                   & 9.2\\ 
77      & 4799.01       & 65805840      &  $ 4.023463 \pm 2.5\cdot 10^{-5} $               &  $ 0.134 \pm 0.013 $                          &  $ 7.5600 \pm 0.0098 $                            & 59    & 7     & 10.8                  & 8.3\\ 
78      & 4825.01       & 53047383      &  $ 2.81847 \pm 0.00025 $                       &  $ 0.1572 \pm 0.0094 $                        &  $ 1.72000 \pm 0.00093 $                           & 45    & 4     & 6.4                   & 8.2\\ 
79      & 4867.01       & 74484423      &  $ 2.835273 \pm 2.1\cdot 10^{-5} $               &  $ 0.177 \pm 0.037 $                          &  $ 3.3700 \pm 0.0053 $                            & 52    & 5     & 8.9                   & 4.6\\ 
80      & 4877.01       & 447054140     &  $ 5.222053 \pm 1.7\cdot 10^{-5} $               &  $ 0.1768 \pm 0.0090 $                        &  $ 4.8500 \pm 0.0014 $                            & 58    & 4     & 6.6                   & 8.5\\ 
81      & 4885.01       & 72504752      &  $ 6.939285 \pm 3.6\cdot 10^{-5} $               &  $ 0.244 \pm 0.011 $                          &  $ 14.5900 \pm 0.0066 $                            & 51    & 3     & 8.0                   & 35.7\\ 
82      & 4936.01       & 68030406      &  $ 4.775018 \pm 2.0\cdot 10^{-5} $               &  $ 0.1517 \pm 0.0079 $                        &  $ 9.4700 \pm 0.0054 $                            & 84    & 4     & 9.4                   & 13.3\\ 
83      & 4953.01       & 405571798     &  $ 2.959018 \pm 1.8\cdot 10^{-5} $               &  $ 0.1180 \pm 0.0086 $                        &  $ 12.520 \pm 0.016 $                             & 64    & 5     & 15.4                  & 14.0\\ 
84      & 4968.01       & 253961767     &  $ 2.23481 \pm 0.00048 $                       &  $ 0.1385 \pm 0.0051 $                        &  $ 7.2600 \pm 0.0055 $                            & 78    & 6     & 9.9                   & 12.8\\ 
85      & 4969.01       & 406106335     &  $ 3.077366 \pm 2.2\cdot 10^{-5} $               &  $ 0.1491 \pm 0.0071 $                        &  $ 5.1400 \pm 0.0048 $                            & 64    & 3     & 5.1                   & 10.5\\ 
86      & 4991.01       & 247180220     &  $ 0.77021 \pm 0.00020 $                       &  $ 0.036 \pm 0.011 $                          &  $ 29.13 \pm 0.52 $                              & 44    & 2     & 8.7                   & 69.0\\ 
87      & 4992.01       & 362111691     &  $ 4.880747 \pm 2.0\cdot 10^{-5} $               &  $ 0.1312 \pm 0.0098 $                        &  $ 11.4400 \pm 0.0088 $                            & 71    & 2     & 11.4                  & 17.2\\ 
88      & 4996.01       & 384709136     &  $ 4.813141 \pm 2.4\cdot 10^{-5} $               &  $ 0.206 \pm 0.012 $                          &  $ 5.0300 \pm 0.0024 $                            & 52    & 4     & 4.6                   & 4.7\\ 
89      & 5002.01       & 418471116     &  $ 3.731882 \pm 1.3\cdot 10^{-5} $               &  $ 0.1346 \pm 0.0073 $                        &  $ 10.8500 \pm 0.0039 $                            & 41    & 4     & 5.9                   & 13.2\\ 
90      & 5021.01       & 403405969     &  $ 12.97506 \pm 0.00018 $                      &  $ 0.255 \pm 0.030 $                          &  $ 6.210 \pm 0.012 $                             & 46    & 2     & 4.7                   & 6.7\\ 
91      & 5050.01       & 282701278     &  $ 2.306457 \pm 1.1\cdot 10^{-5} $               &  $ 0.0863 \pm 0.0082 $                        &  $ 5.6500 \pm 0.0075 $                            & 45    & 3     & 4.0                   & 6.3\\ 
92      & 5052.01       & 208151094     &  $ 4.06681 \pm 0.00077 $                       &  $ 0.1245 \pm 0.0074 $                        &  $ 19.690 \pm 0.010 $                             & 50    & 2     & 10.8                  & 20.0\\ 
93      & 5054.01       & 401704685     &  $ 4.210006 \pm 2.6\cdot 10^{-5} $               &  $ 0.111 \pm 0.014 $                          &  $ 7.9900 \pm 0.0041 $                            & 46    & 2     & 6.2                   & 9.8\\ 
94      & 5061.01       & 214218272     &  $ 3.364535 \pm 2.3\cdot 10^{-5} $               &  $ 0.131 \pm 0.013 $                          &  $ 7.010 \pm 0.010 $                             & 49    & 3     & 10.4                  & 12.0\\ 
95      & 5073.01       & 458686969     &  $ 3.1393697 \pm 2.8\cdot 10^{-6} $               &  $ 0.1656 \pm 0.0018 $                        &  $ 9.09000 \pm 0.00031 $                           & 63    & 2     & 8.5                   & 10.8\\ 
\hline
\multicolumn{10}{| l |}{\textbf{Transit confirmations in binary systems}}\\
\hline 
1       & 3722.01       & 346015394     &  $ 1.078948 \pm 6.2\cdot 10^{-5} $               &  $ 0.0770 \pm 0.0053 $                        &  $ 13.4100 \pm 0.0070 $                            & 60    & 6     & 11.9                  & 30.7  \\ 
 \end{longtable}
\begin{flushleft} 
\qquad$^{b}$ As described in Sect.~\ref{sec:transit_depths}, we adopt a uniform value of $\sigma_\mathrm{d} = 6.5$ mmag for the Gaia transit depth uncertainty of on-target confirmations.\\
\end{flushleft}
}
 
 
 {\tiny \tabcolsep=7pt 
\LTcapwidth=0.95\textwidth
\begin{longtable}[]{|lrrccc | rrrc|}
\caption{BEBs in Phase~II.} 
\label{tab:group_params_bebs}\\
 \hline
 \multicolumn{6}{| l |}{\textbf{TESS}} & \multicolumn{4}{| l |}{\textbf{Gaia}}\\
 \hline
  & \multicolumn{1}{ c }{TOI} & \multicolumn{1}{ c }{TIC} & Orbital Period & Transit Duration & Transit Depth &  \multicolumn{1}{ c }{$N$}  & \multicolumn{1}{ c }{$\smallsub{N}{IT}$} & $\mathcal{S/N}_\mathrm{T}$ & Transit Depth \\
    &  &     & [day]            & [day]              & [ppt]          &     &      &      &   [ppt] \\
 \hline
 \endfirsthead

 \hline
 \multicolumn{10}{|c|}{Table \ref{tab:group_params_bebs} continued.}\\
 \hline
 \multicolumn{6}{| l |}{\textbf{TESS}} & \multicolumn{4}{| l |}{\textbf{Gaia}}\\
 \hline
  & \multicolumn{1}{ c }{TOI} & \multicolumn{1}{ c }{TIC} & Orbital Period & Transit Duration & Transit Depth &  \multicolumn{1}{ c }{$N$}  & \multicolumn{1}{ c }{$\smallsub{N}{IT}$} & $\mathcal{S/N}_\mathrm{T}$ & Transit Depth \\
    &  &     & [day]            & [day]              & [ppt]          &     &      &      &   [ppt] \\
 \hline
 \endhead

 \hline
 \endfoot

 \hline
 \endlastfoot
1       & 3507.01       & 86451883      &  $ 1.7561 \pm 0.0010 $                         &  $ 0.119 \pm 0.022 $                          &  $ 11.010 \pm 0.048 $                             & 50    & 4     & 17.4                  &  $ 114.4 \pm 7.3 $                      \\ 
2       & 3508.01       & 278968371     &  $ 1.29978 \pm 0.00034 $                       &  $ 0.098 \pm 0.011 $                          &  $ 7.140 \pm 0.012 $                             & 55    & 5     & 31.1                  &  $ 101.7 \pm 3.6 $                      \\ 
3       & 3525.01       & 373576942     &  $ 0.89122 \pm 0.00031 $                       &  $ 0.077 \pm 0.013 $                          &  $ 9.160 \pm 0.024 $                             & 60    & 4     & 20.4                  &  $ 172.3 \pm 9.5 $                      \\ 
4       & 3526.01       & 89876424      &  $ 1.53807 \pm 0.00034 $                       &  $ 0.093 \pm 0.013 $                          &  $ 8.770 \pm 0.025 $                             & 43    & 2     & 11.6                  &  $ 198 \pm 19 $                         \\ 
5       & 3541.01       & 291147951     &  $ 2.53195 \pm 0.00041 $                       &  $ 0.107 \pm 0.015 $                          &  $ 8.850 \pm 0.018 $                             & 41    & 3     & 13.1                  &  $ 43.3 \pm 3.5 $                       \\ 
6       & 3548.01       & 271167979     &  $ 0.309395 \pm 1.0\cdot 10^{-5} $               &  $ 0.0184 \pm 0.0035 $                        &  $ 7.7 \pm 4.0 $                               & 36    & 2     & 15.1                  &  $ 210 \pm 16 $                         \\ 
7       & 3562.01       & 193708920     &  $ 1.32985 \pm 0.00030 $                       &  $ 0.1253 \pm 0.0058 $                        &  $ 2.19 \pm 0.80 $                              & 39    & 7     & 23.0                  &  $ 113.0 \pm 5.8 $                      \\ 
8       & 3566.01       & 169187982     &  $ 1.76444 \pm 0.00019 $                       &  $ 0.090 \pm 0.020 $                          &  $ 20.55 \pm 0.43 $                              & 44    & 2     & 24.0                  &  $ 225 \pm 11 $                         \\ 
9       & 3578.01       & 358186451     &  $ 1.31707 \pm 0.00029 $                       &  $ 0.073 \pm 0.016 $                          &  $ 0.6400 \pm 0.0028 $                            & 45    & 3     & 15.7                  &  $ 63.4 \pm 4.3 $                       \\ 
10      & 3579.01       & 316606244     &  $ 3.3106 \pm 0.0014 $                         &  $ 0.162 \pm 0.034 $                          &  $ 0.5900 \pm 0.0026 $                            & 61    & 3     & 18.2                  &  $ 82.1 \pm 4.8 $                       \\ 
11      & 3580.01       & 305478010     &  $ 3.12134 \pm 0.00019 $                       &  $ 0.1805 \pm 0.0072 $                        &  $ 26.3000 \pm 0.0066 $                            & 74    & 2     & 13.3                  &  $ 32.9 \pm 2.6 $                       \\ 
12      & 3581.01       & 295413003     &  $ 3.03730 \pm 0.00036 $                       &  $ 0.194 \pm 0.026 $                          &  $ 6.29 \pm 0.57 $                              & 40    & 2     & 26.5                  &  $ 119.1 \pm 4.9 $                      \\ 
13      & 3608.01       & 267867517     &  $ 2.68163 \pm 0.00072 $                       &  $ 0.112 \pm 0.025 $                          &  $ 3.100 \pm 0.017 $                             & 47    & 2     & 15.0                  &  $ 132.9 \pm 9.7 $                      \\ 
14      & 3627.01       & 1979500576    &  $ 2.63427 \pm 0.00049 $                       &  $ 0.167 \pm 0.022 $                          &  $ 3.19 \pm 0.57 $                              & 49    & 4     & 26.3                  &  $ 198.5 \pm 8.7 $                      \\ 
15      & 3659.01       & 374778982     &  $ 2.3878 \pm 0.0021 $                         &  $ 0.108 \pm 0.026 $                          &  $ 6.800 \pm 0.045 $                             & 76    & 3     & 9.0                   &  $ 38.3 \pm 4.4 $                       \\ 
16      & 3672.01       & 50943163      &  $ 2.4254 \pm 0.0013 $                         &  $ 0.138 \pm 0.022 $                          &  $ 1.9100 \pm 0.0051 $                            & 57    & 4     & 56.1                  &  $ 137.6 \pm 2.7 $                      \\ 
17      & 3745.01       & 666567477     &  $ 1.68831 \pm 0.00079 $                       &  $ 0.111 \pm 0.015 $                          &  $ 7.400 \pm 0.017 $                             & 34    & 3     & 6.8                   &  $ 42.2 \pm 6.6 $                       \\ 
18      & 3746.01       & 284747017     &  $ 1.80127 \pm 0.00093 $                       &  $ 0.133 \pm 0.018 $                          &  $ 1.8000 \pm 0.0036 $                            & 47    & 3     & 14.1                  &  $ 117.1 \pm 9.1 $                      \\ 
19      & 3781.01       & 470974162     &  $ 3.25885 \pm 0.00084 $                       &  $ 0.094 \pm 0.027 $                          &  $ 4.580 \pm 0.022 $                             & 35    & 2     & 12.1                  &  $ 62.0 \pm 5.4 $                       \\ 
20      & 3924.01       & 456042145     &  $ 2.047483 \pm 6.4\cdot 10^{-5} $               &  $ 0.132 \pm 0.012 $                          &  $ 6.8 \pm 2.4 $                               & 44    & 7     & 52.4                  &  $ 205.0 \pm 4.8 $                      \\ 
21      & 3925.01       & 435167167     &  $ 3.03309 \pm 0.00018 $                       &  $ 0.168 \pm 0.027 $                          &  $ 1.2800 \pm 0.0036 $                            & 38    & 2     & 12.2                  &  $ 155 \pm 14 $                         \\ 
22      & 3979.01       & 444523623     &  $ 2.04166 \pm 0.00012 $                       &  $ 0.147 \pm 0.040 $                          &  $ 3.230 \pm 0.017 $                             & 60    & 3     & 25.3                  &  $ 179.1 \pm 8.0 $                      \\ 
23      & 3981.01       & 312548415     &  $ 2.24375 \pm 0.00014 $                       &  $ 0.159 \pm 0.027 $                          &  $ 2.8800 \pm 0.0087 $                            & 67    & 5     & 15.0                  &  $ 49.2 \pm 3.5 $                       \\ 
24      & 3985.01       & 604863428     &  $ 3.23585 \pm 0.00017 $                       &  $ 0.157 \pm 0.025 $                          &  $ 5.460 \pm 0.016 $                             & 56    & 2     & 9.9                   &  $ 33.6 \pm 3.5 $                       \\ 
25      & 3990.01       & 420446142     &  $ 5.84236 \pm 0.00022 $                       &  $ 0.101 \pm 0.017 $                          &  $ 2.2800 \pm 0.0070 $                            & 55    & 2     & 41.2                  &  $ 144.0 \pm 3.8 $                      \\ 
26      & 3991.01       & 418133917     &  $ 2.27748 \pm 0.00012 $                       &  $ 0.156 \pm 0.025 $                          &  $ 1.5200 \pm 0.0040 $                            & 53    & 4     & 17.2                  &  $ 34.1 \pm 2.1 $                       \\ 
27      & 3994.01       & 372505528     &  $ 3.897414 \pm 7.8\cdot 10^{-5} $               &  $ 0.086 \pm 0.016 $                          &  $ 1.9900 \pm 0.0032 $                            & 49    & 3     & 8.9                   &  $ 30.2 \pm 3.6 $                       \\ 
28      & 4005.01       & 395773226     &  $ 3.84717 \pm 0.00011 $                       &  $ 0.092 \pm 0.018 $                          &  $ 1.7700 \pm 0.0064 $                            & 42    & 2     & 22.6                  &  $ 206 \pm 10 $                         \\ 
29      & 4036.01       & 367913453     &  $ 2.77616 \pm 0.00013 $                       &  $ 0.089 \pm 0.021 $                          &  $ 1.3400 \pm 0.0068 $                            & 44    & 3     & 16.2                  &  $ 44.8 \pm 2.9 $                       \\ 
30      & 4048.01       & 279605855     &  $ 1.972401 \pm 3.0\cdot 10^{-5} $               &  $ 0.101 \pm 0.026 $                          &  $ 7.460 \pm 0.020 $                             & 42    & 2     & 22.0                  &  $ 332 \pm 19 $                         \\ 
31      & 4052.01       & 236775063     &  $ 0.7409429 \pm 7.0\cdot 10^{-6} $               &  $ 0.045 \pm 0.010 $                          &  $ 2.4600 \pm 0.0070 $                            & 38    & 3     & 32.3                  &  $ 390 \pm 16 $                         \\ 
32      & 4068.01       & 159334269     &  $ 1.1578392 \pm 7.2\cdot 10^{-6} $               &  $ 0.057 \pm 0.014 $                          &  $ 9.660 \pm 0.024 $                             & 33    & 2     & 14.6                  &  $ 157 \pm 12 $                         \\ 
33      & 4083.01       & 399444057     &  $ 2.071480 \pm 5.6\cdot 10^{-5} $               &  $ 0.130 \pm 0.019 $                          &  $ 1.6600 \pm 0.0023 $                            & 60    & 5     & 56.0                  &  $ 149.6 \pm 3.0 $                      \\ 
34      & 4126.01       & 141616639     &  $ 3.24287 \pm 0.00015 $                       &  $ 0.101 \pm 0.021 $                          &  $ 1.4700 \pm 0.0066 $                            & 49    & 2     & 17.1                  &  $ 95.5 \pm 6.0 $                       \\ 
35      & 4130.01       & 71396541      &  $ 6.26972 \pm 0.00027 $                       &  $ 0.154 \pm 0.032 $                          &  $ 1.2400 \pm 0.0033 $                            & 75    & 3     & 24.0                  &  $ 40.1 \pm 1.7 $                       \\ 
36      & 4134.01       & 420803693     &  $ 0.913597 \pm 2.4\cdot 10^{-5} $               &  $ 0.052 \pm 0.014 $                          &  $ 0.5400 \pm 0.0032 $                            & 47    & 2     & 7.0                   &  $ 133 \pm 21 $                         \\ 
37      & 4135.01       & 288471040     &  $ 2.513920 \pm 3.3\cdot 10^{-5} $               &  $ 0.0994 \pm 0.0082 $                        &  $ 0.59000 \pm 0.00043 $                           & 33    & 4     & 13.4                  &  $ 14.2 \pm 1.1 $                       \\ 
38      & 4139.01       & 408729765     &  $ 4.62008 \pm 0.00011 $                       &  $ 0.090 \pm 0.015 $                          &  $ 3.4800 \pm 0.0090 $                            & 48    & 3     & 38.5                  &  $ 209.1 \pm 6.3 $                      \\ 
39      & 4143.01       & 371711170     &  $ 9.84822 \pm 0.00073 $                       &  $ 0.177 \pm 0.034 $                          &  $ 1.6700 \pm 0.0062 $                            & 48    & 2     & 18.4                  &  $ 100.0 \pm 5.8 $                      \\ 
40      & 4164.01       & 420108587     &  $ 3.52313 \pm 0.00012 $                       &  $ 0.110 \pm 0.018 $                          &  $ 0.8200 \pm 0.0022 $                            & 29    & 5     & 11.6                  &  $ 12.9 \pm 1.2 $                       \\ 
41      & 4211.01       & 385654601     &  $ 2.2715616 \pm 8.4\cdot 10^{-6} $               &  $ 0.114 \pm 0.010 $                          &  $ 8.1900 \pm 0.0066 $                            & 35    & 4     & 7.2                   &  $ 19.0 \pm 2.9 $                       \\ 
42      & 4216.01       & 187861378     &  $ 2.181681 \pm 1.4\cdot 10^{-5} $               &  $ 0.160 \pm 0.020 $                          &  $ 1.2800 \pm 0.0017 $                            & 56    & 3     & 15.2                  &  $ 45.5 \pm 3.1 $                       \\ 
43      & 4225.01       & 141405814     &  $ 1.1355486 \pm 8.4\cdot 10^{-6} $               &  $ 0.110 \pm 0.014 $                          &  $ 0.39000 \pm 0.00074 $                           & 43    & 3     & 44.5                  &  $ 163.1 \pm 4.1 $                      \\ 
44      & 4229.01       & 319282294     &  $ 1.13857 \pm 0.00046 $                       &  $ 0.079 \pm 0.014 $                          &  $ 6.220 \pm 0.024 $                             & 50    & 4     & 29.7                  &  $ 205.2 \pm 8.0 $                      \\ 
45      & 4232.01       & 126357909     &  $ 1.22480 \pm 0.00053 $                       &  $ 0.069 \pm 0.014 $                          &  $ 7.640 \pm 0.037 $                             & 53    & 5     & 17.8                  &  $ 123.4 \pm 7.8 $                      \\ 
46      & 4236.01       & 36227841      &  $ 0.4021282 \pm 1.9\cdot 10^{-6} $               &  $ 0.034 \pm 0.012 $                          &  $ 2.4 \pm 1.0 $                               & 58    & 6     & 10.0                  &  $ 6.64 \pm 0.71 $                      \\ 
47      & 4243.01       & 400683848     &  $ 1.0883300 \pm 8.6\cdot 10^{-6} $               &  $ 0.101 \pm 0.020 $                          &  $ 0.5400 \pm 0.0016 $                            & 48    & 6     & 10.0                  &  $ 61.3 \pm 6.8 $                       \\ 
48      & 4256.01       & 77102895      &  $ 1.0962372 \pm 6.5\cdot 10^{-6} $               &  $ 0.097 \pm 0.017 $                          &  $ 2.4600 \pm 0.0042 $                            & 48    & 4     & 30.0                  &  $ 189.4 \pm 7.3 $                      \\ 
49      & 4258.01       & 74484886      &  $ 3.850566 \pm 2.7\cdot 10^{-5} $               &  $ 0.104 \pm 0.014 $                          &  $ 6.070 \pm 0.012 $                             & 46    & 4     & 48.5                  &  $ 211.4 \pm 5.1 $                      \\ 
50      & 4292.01       & 255918903     &  $ 1.72035 \pm 0.00018 $                       &  $ 0.1086 \pm 0.0087 $                        &  $ 7.4900 \pm 0.0064 $                            & 72    & 7     & 8.8                   &  $ 20.7 \pm 2.5 $                       \\ 
51      & 4300.01       & 266843799     &  $ 0.8062768 \pm 2.8\cdot 10^{-6} $               &  $ 0.0818 \pm 0.0095 $                        &  $ 0.66000 \pm 0.00082 $                           & 55    & 2     & 19.5                  &  $ 286 \pm 18 $                         \\ 
52      & 4335.01       & 443658196     &  $ 4.4990 \pm 0.0024 $                         &  $ 0.257 \pm 0.021 $                          &  $ 4.08 \pm 0.30 $                              & 51    & 2     & 40.7                  &  $ 225.3 \pm 6.4 $                      \\ 
53      & 4346.01       & 389371332     &  $ 3.907712 \pm 4.6\cdot 10^{-5} $               &  $ 0.397 \pm 0.032 $                          &  $ 0.600 \pm 0.048 $                             & 201   & 14    & 44.4                  &  $ 465 \pm 15 $                         \\ 
54      & 4396.01       & 379723086     &  $ 7.439891 \pm 4.4\cdot 10^{-5} $               &  $ 0.120 \pm 0.012 $                          &  $ 1.0100 \pm 0.0014 $                            & 75    & 3     & 27.7                  &  $ 148.2 \pm 5.9 $                      \\ 
55      & 4397.01       & 107728509     &  $ 2.6646 \pm 0.0028 $                         &  $ 0.172 \pm 0.032 $                          &  $ 0.8400 \pm 0.0032 $                            & 36    & 5     & 22.5                  &  $ 264 \pm 15 $                         \\ 
56      & 4423.01       & 344600402     &  $ 1.3875331 \pm 9.7\cdot 10^{-6} $               &  $ 0.099 \pm 0.017 $                          &  $ 1.1400 \pm 0.0021 $                            & 41    & 3     & 24.9                  &  $ 322 \pm 16 $                         \\ 
57      & 4498.01       & 305943739     &  $ 5.30735 \pm 0.00011 $                       &  $ 0.155 \pm 0.029 $                          &  $ 0.20000 \pm 0.00057 $                           & 93    & 3     & 15.0                  &  $ 117.8 \pm 8.4 $                      \\ 
58      & 4502.01       & 239372503     &  $ 1.71043 \pm 0.00076 $                       &  $ 0.053 \pm 0.029 $                          &  $ 0.2500 \pm 0.0050 $                            & 47    & 5     & 14.0                  &  $ 67.6 \pm 5.3 $                       \\ 
59      & 4808.01       & 126772538     &  $ 1.470060 \pm 1.0\cdot 10^{-5} $               &  $ 0.091 \pm 0.017 $                          &  $ 4.760 \pm 0.015 $                             & 58    & 4     & 19.6                  &  $ 190 \pm 11 $                         \\ 
60      & 4812.01       & 81744640      &  $ 3.313845 \pm 4.0\cdot 10^{-5} $               &  $ 0.161 \pm 0.032 $                          &  $ 3.5500 \pm 0.0097 $                            & 71    & 2     & 10.4                  &  $ 30.8 \pm 3.1 $                       \\ 
61      & 4829.01       & 268439377     &  $ 0.9855374 \pm 3.8\cdot 10^{-6} $               &  $ 0.0492 \pm 0.0070 $                        &  $ 3.2900 \pm 0.0074 $                            & 33    & 3     & 23.0                  &  $ 354 \pm 20 $                         \\ 
62      & 4840.01       & 80029842      &  $ 2.399288 \pm 2.6\cdot 10^{-5} $               &  $ 0.106 \pm 0.017 $                          &  $ 1.1800 \pm 0.0040 $                            & 37    & 2     & 13.8                  &  $ 118.4 \pm 9.4 $                      \\ 
63      & 4846.01       & 418971787     &  $ 1.0638184 \pm 7.1\cdot 10^{-6} $               &  $ 0.133 \pm 0.016 $                          &  $ 3.2200 \pm 0.0048 $                            & 46    & 3     & 18.1                  &  $ 208 \pm 13 $                         \\ 
64      & 4870.01       & 446135868     &  $ 1.811917 \pm 1.6\cdot 10^{-5} $               &  $ 0.109 \pm 0.024 $                          &  $ 5.510 \pm 0.020 $                             & 56    & 6     & 19.0                  &  $ 64.0 \pm 3.7 $                       \\ 
65      & 4874.01       & 463754499     &  $ 2.286686 \pm 2.3\cdot 10^{-5} $               &  $ 0.120 \pm 0.016 $                          &  $ 3.7200 \pm 0.0083 $                            & 42    & 3     & 16.9                  &  $ 37.6 \pm 2.3 $                       \\ 
66      & 4876.01       & 458300989     &  $ 4.13795 \pm 0.00016 $                       &  $ 0.197 \pm 0.032 $                          &  $ 2.610 \pm 0.010 $                             & 45    & 6     & 14.3                  &  $ 76.7 \pm 6.0 $                       \\ 
67      & 4894.01       & 469333534     &  $ 3.561122 \pm 4.8\cdot 10^{-5} $               &  $ 0.114 \pm 0.024 $                          &  $ 5.240 \pm 0.023 $                             & 47    & 2     & 9.1                   &  $ 48.1 \pm 5.5 $                       \\ 
68      & 4902.01       & 358072664     &  $ 4.193494 \pm 3.5\cdot 10^{-5} $               &  $ 0.078 \pm 0.014 $                          &  $ 2.7100 \pm 0.0091 $                            & 52    & 4     & 20.2                  &  $ 186 \pm 11 $                         \\ 
69      & 4946.01       & 319783554     &  $ 1.819376 \pm 2.9\cdot 10^{-5} $               &  $ 0.124 \pm 0.042 $                          &  $ 5.240 \pm 0.081 $                             & 60    & 6     & 23.8                  &  $ 94.5 \pm 4.4 $                       \\ 
70      & 4964.01       & 406443564     &  $ 1.1167145 \pm 4.3\cdot 10^{-6} $               &  $ 0.0757 \pm 0.0054 $                        &  $ 12.81 \pm 0.74 $                              & 67    & 3     & 32.5                  &  $ 246.0 \pm 8.8 $                      \\ 
71      & 5006.01       & 449107129     &  $ 4.254955 \pm 2.8\cdot 10^{-5} $               &  $ 0.176 \pm 0.038 $                          &  $ 2.3700 \pm 0.0031 $                            & 43    & 4     & 10.9                  &  $ 21.7 \pm 2.1 $                       \\ 
72      & 5075.01       & 23113034      &  $ 0.856742 \pm 5.7\cdot 10^{-5} $               &  $ 0.0522 \pm 0.0078 $                        &  $ 0.34000 \pm 0.00070 $                           & 48    & 3     & 64.1                  &  $ 315.4 \pm 6.1 $                      \\ 
 \end{longtable}}



\end{appendix}
\end{document}